\documentclass[%
reprint,prl,
superscriptaddress,
amsmath,amssymb,
aps,
]{revtex4-1}

\usepackage{xparse}
\usepackage{braket}

\usepackage{mathtools}
\usepackage{appendix}
\usepackage{graphicx}
\usepackage{dcolumn}
\usepackage{bm}
\usepackage{hyperref}
\usepackage{xcolor}
\usepackage[normalem]{ulem}
\usepackage{cancel}

\usepackage{float}
\usepackage{braket}
\usepackage{bbold}
\usepackage{mathtools}

\DeclareMathOperator{\Tr}{Tr}

\newcommand{\STr}{\mathrm{STr}}

\newcommand{\msf}[1]{\mathsf{#1}}

\newcommand{\D}{\mathcal{D}}

\newcommand{\e}{\varepsilon}

\newcommand{\w}{\omega}

\newcommand{\TZ}{\tilde{Z}}

\newcommand{\aF}{\alpha_{\mathrm{F}}}
\newcommand{\aB}{\alpha_{\mathrm{B}}}
\newcommand{\bF}{\beta_{\mathrm{F}}}
\newcommand{\bB}{\beta_{\mathrm{B}}}

\newcommand{\re}{\operatorname{Re}}

\begin{document}

\title{
		Field theory approach to eigenstate thermalization in random quantum circuits
	}

\author{Yunxiang Liao}

\affiliation{Joint Quantum Institute, Department of Physics, University of Maryland, College Park, MD 20742, USA.}
\affiliation{Condensed Matter Theory Center, Department of Physics, University of Maryland, College Park, MD 20742, USA.}

\author{Victor Galitski}

\affiliation{Joint Quantum Institute, Department of Physics, University of Maryland, College Park, MD 20742, USA.}

\date{\today}

\begin{abstract}
We use field-theoretic methods to explore the statistics of eigenfunctions of the Floquet operator for a large family of Floquet random quantum circuits. The correlation function of the quasienergy eigenstates is calculated and shown to exhibit random matrix circular unitary ensemble
statistics, which is consistent with the analogue of Berry's conjecture for quantum circuits.  This quantity determines all key metrics of quantum chaos, such as the spectral form factor and thermalizing time-dependence of the expectation value of an arbitrary observable.  It also allows us to explicitly show that the matrix elements of local operators satisfy  the eigenstate thermalization hypothesis (ETH); i.e., the variance of the off-diagonal matrix elements of such operators is exponentially small in the system size.  These results represent a proof of ETH for the family of Floquet random quantum circuits at a physical level of rigor.  An outstanding open question for this and most of other sigma-model calculations is a mathematically rigorous proof of the validity of the saddle-point approximation in the large-$N$ limit.
\end{abstract}
\maketitle

Quantum statistical mechanics is a cornerstone of modern physics.  However, understanding of its microscopic origins is still incomplete.  How does an isolated quantum many-body system, prepared in a generic initial state, reach thermal equilibrium under its own unitary dynamics?
It was proposed by Deutsch~\cite{Deutsch} and Srednicki~\cite{srednicki1994,srednicki1999} that the emergence of the statistical-mechanical behavior in isolated quantum many-body systems can be understood based on the eigenstate thermalization hypothesis (ETH).  It states that the matrix element of a physical observable $O$ in the energy eigenbasis of a thermalizing system has the form~\cite{srednicki1999}
\begin{align}\label{eq:ETH}
\begin{aligned}
O_{\mu\nu}
=
\bar{O}(\bar{E}_{\mu\nu}) \delta_{\mu\nu}
+
\Omega^{-1/2}(\bar{E}_{\mu\nu})
f(\bar{E}_{\mu\nu},\Delta E_{\mu\nu})R_{\mu\nu}.
\end{aligned}
\end{align}
Here $O_{\mu\nu}=\bra{\mu} O \ket{\nu}$, where $\ket{\nu}$ represents the energy eigenstate with energy $E_{\nu}$. $\bar{E}_{\mu\nu}=(E_{\mu}+E_{\nu})/2$ and $\Delta E_{\mu\nu}=E_{\mu}-E_{\nu}$.
$R_{\mu\nu}$  is a  random variable with zero mean and unit variance.
$\Omega(\bar{E})$ is the many-body density of states at energy $\bar{E}$.
$\bar{O}(\bar{E})$ and $f(\bar{E},\Delta E)$ are smooth functions of their arguments, and $f(\bar{E},\Delta E)$ is of order unity.
From the ETH ansatz, it follows that for an arbitrary initial state with a small quantum uncertainty in total energy,  the expectation value of the observable $O$ reaches its equilibrium value at long time with temporal fluctuations exponentially small in system size.

	ETH is closely related to the random matrix theory (RMT)~\cite{Wigner_1955,Wigner_1957,Wigner_1958,Dyson-I,Dyson1962,Mehta,Haake}, and connects thermalization with the notion of quantum chaos~\cite{ETH-review,Deutsch_rev}.
	For a quantum system whose classical limit is chaotic, it was conjectured that the statistics of its energy levels is the same as that of a random matrix ensemble in a suitable symmetry class~\cite{BGS,BerrySFF}.
The  appearance of the universal  RMT level statistics in a quantum system has been widely used as a definition of quantum chaos and countless efforts have been undertaken to understand the origin of this universality~\cite{Sieber_2001, Muller_2005, Prosen-SFF1,Chalker-1,Chalker-2,Chalker-3,Chalker-4, Chalker-5,Chalker-6, Prosen-SFF3,Prosen-SFF2,Prosen-GSFF,Huse2021, chan2021,Prosen-Fermion,Prosen-Boson,Prosen-KIsing,SYK-Altland,SSS, PRB,PRR,PRL,Winer} in a variety of seemingly unrelated  systems.  

Apart from the level statistics another metric of quantum chaos is the statistics of energy eigenstates.  For quantum systems with classically chaotic counterparts Berry's conjecture~\cite{Berry-conj} is expected to apply,  which postulates that an energy eigenstate can be considered as a random superposition of some basis states with Gaussian distributed amplitudes. The ETH follows from Berry's conjecture and connects the statistics of energy eigenstates with the thermalization of physical observables. The ETH has been numerically verified in various systems~\cite{ETH-review, Rigol-ETH-3, Rigol-ETH-0, Rigol-ETH-1, Rigol-ETH-2,Rigol-ETH-4,Rigol-ETH-5,ETH-SpinChain,Huse-ETH, ETH-num-1, ETH-num-2, ETH-Ising,ETH-Ising-2, ETH-Holstein}. However,  its formal proof is still lacking and its range of validity is  unknown.

This Letter explores the validity of an analogue of Berry's conjecture and the ETH	 in random quantum circuits (RQCs), which are ideal platforms to study chaos and thermalization (see reviews Refs.~\cite{Fisher-review,potter} and references therein). 
	It has been demonstrated that RQCs mimic generic many-body quantum chaotic systems
    ~\cite{Nahum-2017,Nahum-2018,Nahum-2018,Keyserlingk-1,Keyserlingk-2,Huse-2018, HJ-2018,Chalker-1,Bertini,Prosen-DUQC-corr,Prosen-DUQC-dyn,Prosen-QDUC-EE-I,Prosen-QDUC-EE-II,Gopalakrishnan}.
	In particular, ETH has been carefully investigated in certain Floquet quantum circuits making use of some of their fine-tuned features~\cite{Prosen-ETH,Chalker-4, Chalker-ETH}.
This work puts forward a field-theoretic approach~\cite{entropy} which allows to calculate  the statistics of quasienergy eigenstates and ETH in a large family of Floquet RQCs.
		
Consider the correlation function of (quasi)energy eigenstates ($\left| \nu\right\rangle$,$\left| \mu\right\rangle$) in a certain basis($\left\lbrace\left|n \right\rangle\right\rbrace$)~\cite{Mirlin}:
	\begin{align}\label{eq:Cw}
	\begin{aligned}
	C_{nn'm'm}(\w)
	=\,&
	\left\langle 
	\sum_{\nu,\mu=1}^{N}
	\left\langle n| \nu\right\rangle 
	\left\langle \nu |n'\right\rangle
	\left\langle m'|\mu\right\rangle
	\left\langle \mu|m \right\rangle
	\right. 
	\\
	&\,\,\,\,\,\,\,\,\,\,\,\,\,\,\,\times \left. 
	\delta\left( \w- E_\nu+E_{\mu}\right) 
	\phantom{\sum\limits_1}\!\!\!\!\!\!\!\!\right\rangle.
	\end{aligned}
	\end{align}
	Here the angular bracket represents the averaging over an ensemble of statistically similar systems and $N$ is the Hilbert space dimension.
	Note that we sum over all pairs of (quasi)energy eigenstates ($\ket{\nu}, \ket{\mu}$) whose (quasi)energy separation $\Delta E_{\nu \mu}$ is $\w$~\footnote{For Floquet system, $\delta(\e)$ in Eq.~\ref{eq:Cw} represents the $2\pi$-periodic delta function  $\delta(\e)=\sum_{k=-\infty}^{\infty}e^{ik\w}/2\pi$.}.
$C_{nn'm'm}(\w)$ determines many important physical properties of the system,
including:
	the spectral form factor (SFF)~\cite{BerrySFF} which diagnoses the statistics of the (quasi)energy spectra;
	the partial spectral form factor (PSFF)~\cite{Zoller,Chalker-4} which probes the statistics of both (quasi)energy eigenstates and eigenvalues;
	and the correlation function of operators which connects directly with ETH~\cite{Kurchan,srednicki1999,ETH-Bound,Prosen-ETH,Chalker-ETH}.

	For an ensemble of Floquet (alternatively, Hamiltonian) systems whose time evolution is generated by unitary Floquet operators $U$ from the circular unitary ensemble (CUE) (alternatively, by Hermitian Hamiltonians $H$ from the Gaussian unitary ensemble (GUE))~\cite{Mehta,Haake}, that is the time evolution operator $U(t)=U^t$ and $U\in$ CUE (
 $U(t)=e^{-iHt}$ and $H\in$ GUE),
	one can prove that 
	\begin{align}\label{eq:Cw-exact}
	\begin{aligned}
	C_{nn'm'm} (\w)
	=&
	\frac{1}{N^2-1} 
	 R_2(\w)
	\left(
	\delta_{nn'}\delta_{mm'}
	-
	\frac{1}{N} \delta_{mn} \delta_{n'm'}
	\right)
	\\
&+
\frac{\delta(\w)}{N^2-1}  \left( 
	N \delta_{mn} \delta_{n'm'}
	-
	\delta_{nn'}\delta_{mm'}
	\right),
	\end{aligned}
	\end{align}
	where the two-level correlation function~\cite{Mehta,Haake} $R_2(\w)$ is 
	\begin{align}
	R_2(\w)
	=
	\sum_{\nu,\mu=1}^{N}
	\left\langle 
	\delta\left( \w- E_\nu+E_{\mu}\right) 
	\right\rangle.
	\end{align}
	The derivation of  Eq.~\ref{eq:Cw-exact} can be found in the Supplemental Material~\cite{Sup} (see also Ref.~\cite{Cotler2017, Cotler2020}), and it makes use of the fact that the (quasi)energy eigenvalues and eigenfunctions of the random matrix ensembles are statistically independent. 
	It is expected that the correlation function $C_{nn'm'm}(\w)$ for generic quantum many-body chaotic systems follows the random matrix prediction (Eq.~\ref{eq:Cw-exact} for the unitary symmetry class), 
 when the energy separation $\w$ is much smaller than the Thouless energy~\cite{Cotler2020}.
    See for example Ref.~\cite{Altland-OS} for a sigma-model derivation of  $C_{nn'm'm}(\w)$ for the Sachdev-Ye-Kitaev model. 

The correlation function Eq.~\ref{eq:Cw-exact} also determines thermalizing time-dependence of the expectation value of observables.  Given an initial density matrix $\rho(0)$,  its time evolution is given by $\rho(t) = U(t) \rho(0) U^\dagger(t)$.  
From Eq.~\ref{eq:Cw-exact},   
we find via a  simple calculation
(see the Supplemental Material~\cite{Sup})
\begin{align}\label{eq:rho-CUE}
\begin{aligned}
    \left\langle 
    \rho_{nm}(t)
    \right\rangle
    =
    (\rho_{th})_{nm}
    \frac{N^2-K(t)}{N^2-1}
    +
    \rho_{nm}(0)
    \frac{K(t)-1}{N^2-1},
\end{aligned}
\end{align}
where $\rho_{th}=\mathbb{1}/N$ is the thermal density matrix. $K(t) = \left\langle \left| \Tr U(t) \right|^2 \right\rangle$ is the spectral form factor that starts off at $K(0) = N^2$, but quickly drops to smaller values via an initial slope and then approaches a plateau at $K(t \to \infty) = N$ via a linear-in-$t$ ramp (for unitary symmetry class)~\cite{Haake,Mehta}. This result suggests that in chaotic systems the initial thermalization dynamics happens very quickly (compared to the Heisenberg/plateau time) but this behavior is non-universal and is determined by the Fourier transform of the density of states~\cite{SFF-thermal}. The universal features in thermalization are more subtle and appear at long times in temporal fluctuations around the thermal average, the system eventually settles into. 
	
To connect the correlation function $C_{nn'm'm}(\w)$ and ETH~\cite{Kurchan,Chalker-ETH,Prosen-ETH}, consider the variance of the matrix elements of a Hermitian operator $O$ in the (quasi)energy eigenbasis $\left\langle |O_{\mu\nu}|^2\right\rangle $ averaged over all pairs of $\mu, \nu$ with the energy separation $\Delta E_{\nu\mu}=\w$:
	\begin{align}\label{eq:CO}
	\begin{aligned}
	C_{O}(\w)
	=
	\frac{1}{N}\sum_{\nu,\mu=1}^{N}
	\left\langle 
	|O_{\mu\nu}|^2\delta(\w-E_{\nu}+E_{\mu})\right\rangle,
	\end{aligned}
	\end{align}
	which is the Fourier transform of the autocorrelation function of the operator $O$: $C_{O}(t)=\left\langle \Tr \left( O(t)O \right) \right\rangle/N $. 
 
	For Floquet  many-body chaotic systems which are expected to thermalize to an infinite temperature state~\cite{Rigol-Floquet,Abanin-Floquet}, the corresponding ETH ansatz Eq.~\ref{eq:ETH} acquires the following form~\cite{Chalker-ETH, Prosen-ETH}
	\begin{align}\label{eq:ETH-1}
	\begin{aligned}
		O_{\mu\nu}
		=
		\bar{O}  \delta_{\mu\nu}
		+
		N^{-1/2}
		f(\Delta E_{\mu\nu})R_{\mu\nu}.
	\end{aligned}
	\end{align}
	Here the infinite-temperature thermal average $ \bar{O}=\Tr (O)/N$, and
	the dependence on average quasienergy $\bar{E}_{\mu\nu}$ has been removed~\cite{Moessner-Floquet}.
    For simplicity, we consider a traceless operator $\Tr (O)=0$.
	Substituting Eq.~\ref{eq:ETH-1} into Eq.~\ref{eq:CO}, one obtains
	\begin{align}\label{eq:CO-2}
	\begin{aligned}
	C_{O}(\w)
	=
	\frac{1}{N^2}
	\left\langle \sum_{\nu,\mu=1}^{N}
	|f(\w)|^2 \delta(\w-E_{\nu}+E_{\mu})
	\right\rangle.
	\end{aligned}
	\end{align}
	On the other hand, if $C_{nn'm'm}(\w)$ coincides with the RMT prediction Eq.~\ref{eq:Cw-exact},  we obtain~\cite{Sup}
	\begin{align}\label{eq:CO-1}
	\begin{aligned}
	C_{O}(\w)
    \approx 
    R_2(\w)
	\Tr(O^2)/N^3,
	\end{aligned}
	\end{align}
    in the large $N \rightarrow \infty$ limit.
	Comparison of Eq.~\ref{eq:CO-2} and Eq.~\ref{eq:CO-1} leads to $f(\w) = \sqrt{\Tr(O^2)/N}$.  
    For a few-body observable $O$, $f(\w) $ is therefore of order unity.
    In summary, if the correlation function $C_{nn'm'm}(\w)$ agrees with the RMT prediction, then the variance of the off-diagonal matrix element of the physical operator $O$ in the quasienergy eigenbasis $O_{\mu\nu}$ is of the order $O(N^{-1})$,
	which is consistent with ETH and ensures  small temporal fluctuations of the expectation value of $O$ around its equilibrium value.
	
	In fact, $f(\w) =\sqrt{\Tr(O^2)/N}$, deduced from the CUE result for $C_{nn'm'm}(\w)$ (Eq.~\ref{eq:Cw-exact}), is in agreement with the RMT prediction for the statistics of the matrix element $O_{\mu\nu}$. 
	Assuming the statistics of the (quasi)energy eigenstates is identical to that of the GUE (or CUE), one can show that, to the leading order in $1/N$, the matrix element of any Hermitian operator $O$ can be expressed as~\cite{Sup}, 
	\begin{align}\label{eq:RMT}
	\begin{aligned}
	O_{\mu\nu}
	=
	\frac{\Tr (O)}{N}\delta_{\mu\nu}
	+
       \left(\frac{\Tr (O^2)}{N^2}-\frac{\Tr^2(O)}{N^3}\right)^{1/2}
	R_{\mu\nu},
	\end{aligned}
	\end{align}
    For a traceless operator $O$, this expression is consistent with the one in Ref.~\cite{ETH-review}, and leads to $f(\w)= \sqrt{\Tr(O^2)/N}$.
    The ETH ansatz for generic ergodic systems without any antiunitary symmetry is expected to reduce to this RMT form Eq.~\ref{eq:RMT}~\cite{ETH-review,ETH-RMT} for small enough energy separation $|\Delta E_{\mu\nu}|$.
    The fact that Eq.~\ref{eq:Cw-exact} (or equivalently Eq.~\ref{eq:CO-1}) is consistent with Eq.~\ref{eq:RMT} is not surprising since they are both derived from the RMT eigenstate statistics.
 
Now we prove that for a large family of Floquet random quantum circuits, the correlation function $C_{nn'm'm}(\w)$ is given by the CUE result in Eq.~\ref{eq:Cw-exact}.
This in turn implies that the variance of the off-diagonal matrix element $O_{\mu\nu}$ for these RQCs is consistent with ETH.
We consider RQCs~\cite{entropy}, which consist of an arbitrary dimensional simple square lattice of qudits, evolving under periodic applications of local quantum gates. 
    Specifically, random unitary matrices, drawn independently from the CUE of dimension $q^2$, are applied to different pairs of neighboring qudits at various substeps within one period. We focus on the limit where the single-particle Hilbert space dimension of each qudit $q\rightarrow \infty$. 
    
	In Ref.~\cite{Chalker-1}, Chan et al considered the special case of 1D brickwork model, and studied the SFF, the bipartite entanglement, as well as the two-point and out-of-time-order correlators of the local observables (see also Ref.~\cite{Chalker-4} for finite $q$ case).
	Later in Ref.~\cite{entropy},  we considered more general models and employed a sigma model approach~\cite{Altland-rev} to prove that the statistics of the quasienergy spectrum of any model in this  family of Floquet RQCs agrees with the CUE result, irrespective of the dimensionality of the qudit lattice, the choice of the boundary condition (periodic or open) and the ordering of the local quantum gates.  We are going to employ a similar sigma model to show that $C_{nn'm'm}(\w)$ of these models also follows the CUE result Eq.~\ref{eq:Cw-exact}.

	 For an arbitrary ensemble of Floquet systems, the correlation function $C_{nn'm'm}(\w)$ (Eq.~\ref{eq:Cw}) can be obtained from the generating function
	\begin{align}
	\begin{aligned}
	\mathcal{Z}[J]
	=&
	\left\langle 
	\int_0^{2\pi} \frac{d\phi}{2\pi}
	\dfrac
	{\det \left( 1-\aF e^{i\phi} U+ J^+\right) 
	}
	{\det \left( 1-\aB e^{i\phi} U\right) 
	}
	\right. 
	\\
	&
	\left. 
	\times
		\dfrac
	{
		\det \left( 1-\bF e^{-i\phi} U^{\dagger} +J^-\right) 
	}
	{
		\det \left( 1-\bB e^{-i\phi} U^{\dagger}\right) 
	}
	\right\rangle,
	\end{aligned}
	\end{align}
    where the angular bracket stands for averaging over the ensemble of Floquet operators $U$.
	Taking derivatives of the generating function $\mathcal{Z}[J]$ with respect to the source $J^{\pm}$, we have
    \begin{align}
    \begin{aligned}\label{eq:C2}
	\bar{C}_{nn'm'm}(\aF,\bF)
	=
	\dfrac{\partial^2\mathcal{Z}[J]}{\partial J^+_{n'n}\partial J^-_{mm'}}
	\bigg\lvert_{J=0,\aB=\aF,\bB=\bF}
    \\
    =
    \sum_{k=0}^{\infty}
	\left\langle 
	(U^k)_{nn'}
	(U^{-k})_{m'm}
	\right\rangle
	\left( \aF\bF\right)^k,
	\end{aligned}
	\end{align}
    where we have set $J=0$, $\aB=\aF$ and $\bB=\bF$ in the end
    \footnote{
    As will be apparent later, we find it more convenient for the calculation to set $\aB=\aF$ and $\bB=\bF$ after taking the derivatives.
   $|\aB|, |\bB|$ are chosen to be infinitesimally smaller than $1$ to avoid singularity and to ensure that the bosonic integral for the denominator is well defined.}.
    Here $\left\langle 
	(U^k)_{nn'}
	(U^{-k})_{m'm}
	\right\rangle$ is just the Fourier transform of $C_{nn'm'm}(\w)$ at discrete time $k$ in units of the period. One can therefore see that, when $\aF\bF=e^{i\w}$,
	\begin{align}
	&\begin{aligned}\label{eq:C1}
	C_{nn'm'm}(\w)
	=
	\frac{1}{2\pi}
	\left( 
	\bar{C}_{nn'm'm}
	+
	\bar{C}^{*}_{n'nmm'}
	-
	\delta_{nn'}\delta_{mm'}
	\right).
	\end{aligned}
	\end{align}
	
	Following a standard derivation of the supersymmetric (SUSY) sigma model~\cite{Efetov} for the Floquet systems with the help of the color-flavor transformation~\cite{Zirnbauer_1996,Zirnbauer_1998,Zirnbauer_1999,Zirnbauer_2021}, we find that the generating function $\mathcal{Z}[J]$ can be rewritten as an integral over $2N\times 2N$ supermatrices $Z$ and $\TZ$ (see the Supplementary Material~\cite{Sup}, the review~\cite{Altland-rev} and references therein):
	\begin{align}\label{eq:sigma}
	\begin{aligned}
	&\mathcal{Z}[J]
	=
	\det(1+J^+)\det(1+J^-)
	\left\langle 
	\int D(\tilde{Z},Z)
	e^{-S[\TZ,Z]}
	\right\rangle ,
	\\
	&S[\TZ, Z]
	=
	-
	\STr \ln (1-\tilde{Z} {Z})
	+
	\STr \ln \left( 1- \tilde{Z}\alpha_J \mathbf{U} Z\beta_J \mathbf{U}^{\dagger} \right).
	\end{aligned}
	\end{align}
	Here $Z^{ab}_{ij}$ and $\TZ^{ab}_{ij}$ carry indices $a,b=\msf{B},\msf{F}$ labeling the boson-fermion ($\msf{BF}$) space and indices $i,j=1,2,...,N$ labeling the Hilbert ($\mathcal{H}$) space wherein the Floquet operator acts. They are subject to the constraints $\tilde{Z}^{\msf{FF}}=-Z^{\msf{FF}}\,^{\dagger}$, $\tilde{Z}^{\msf{BB}}=Z^{\msf{BB}}\,^{\dagger}$,  $|\tilde{Z}^{\msf{BB}} Z^{\msf{BB}}|<1$.
    $\STr$ denotes the supertrace over the $\msf{BF}\otimes \mathcal{H}$ space.
	$\alpha_J$ and $\beta_J$ are $2N\times 2N$ supermatrices defined as
	\begin{align}
	\begin{aligned}
	&\alpha_J
	=
	\begin{bmatrix}
	\aB \mathbb{1}_{\mathcal{H}} & 0
	\\
	0 & \aF(\mathbb{1}_{\mathcal{H}}+J^+)^{-1}
	\end{bmatrix}_{\msf{BF}},
	 \\
	 &
	\beta_J
	=
	\begin{bmatrix}
	\bB \mathbb{1}_{\mathcal{H}}  & 0
	\\
	0 & \bF(\mathbb{1}_{\mathcal{H}}+J^-)^{-1}
	\end{bmatrix}_{\msf{BF}},
	\end{aligned}
	\end{align}
    and $\mathbf{U}=U\otimes \mathbb{1}_{\msf{BF}}$.
	We use $\mathbb{1}_{\mathcal{H}}$ ($\mathbb{1}_{\msf{BF}}$) to denote the identity matrix in the Hilbert (boson-fermion) space.
	
	Note that Eq.~\ref{eq:sigma} is exact and applicable to any ensemble of Floquet systems.
	We will now turn to the specific model under consideration here - the family of Floquet RQCs with local Haar-distributed unitary gates, and evaluate the correlation function $C_{nn'm'm}(\w)$ in the large $N$ limit by analyzing the saddle points of the supermatrix $Z$ (Eq.~\ref{eq:sigma}) and the quadratic fluctuations around them. 
    In Ref.~\cite{entropy}, we have proven that, for the current model, the higher order fluctuations give rise to a contribution of higher order in $1/N$ and therefore can be neglected.
	
	Variation of the action $S[\TZ,Z]$ (Eq.~\ref{eq:sigma}) with respect to $\TZ$ yields the saddle point equation:
	\begin{align}\label{eq:SPEQ}
	\begin{aligned}
	Z(1-\tilde{Z} {Z})^{-1}
	=
	\alpha_J \mathbf{U} Z\beta_J  
	\mathbf{U}^{\dagger} 
	\left(1- \tilde{Z} \alpha_J \mathbf{U}  Z \beta_J  \mathbf{U}^{\dagger}\right)^{-1},
	\end{aligned}
	\end{align}
    for an arbitrary configuration of Floquet operator $U$ in the ensemble.
	It is easy to see that the standard saddle point $Z^{(s)}=0$ solves this equation, and the corresponding action vanishes at this saddle point $S^{(s)}=0$.
	Another solution is the nonstandard saddle point of the form~\cite{Andreev-Altshuler,Kamenev,Kamenev-GUE,Kamenev-Keldysh,Altland-rev}
	\begin{align}\label{eq:AA}
	\begin{aligned}
		Z^{(ns)}
		=
		\begin{bmatrix}
		0 & 0
		\\
		0 & z 
		\end{bmatrix}_{\msf{BF}},
		\qquad
		\TZ^{(ns)}
		=
		\begin{bmatrix}
		0 & 0
		\\
		0 & -z^{\dagger}
		\end{bmatrix}_{\msf{BF}}.
	\end{aligned}
	\end{align}
	Here $z$ is a $N\times N$ matrix acting in the Hilbert space, and can be expressed as $z=c_0z'$, with $z'$ being an arbitrary invertiable matrix of order unity and the coefficient $c_0 \rightarrow \infty$.
	The action at the nonstandard saddle point is 
		\begin{align}
		\begin{aligned}\label{eq:Ssaddle-ns}
        S^{(ns)}
		=\, & 
		-\Tr \ln \left[  \alpha_F\beta_F (1+J^+)^{-1}  (1+J^-)^{-1} \right].
		\end{aligned}
		\end{align}
We note that the standard and nonstandard saddle points are solutions to Eq.~\ref{eq:SPEQ} for every configuration of the Floquet operator $U$ in the ensemble. 
There also exist other saddle points 
(given by Eq.~\ref{eq:AA} with singular matrix $z'$) 
which solve the saddle point equation for one or a few specific realizations of the Floquet operator $U$ in the current ensemble.
We believe these saddle points are not important to the ensemble averaged theory, and therefore ignore them in the following calculation.

If one introduces a positive infinitesimal $\eta$ to $\ln(\aF\bF)=i\w-\eta$, the correlation function
$C_{nn'm'm}(\w)$ obtained from Eqs.~\ref{eq:C2} and~\ref{eq:C1} becomes its smoothed version where the $2\pi$-periodic Dirac delta function $\delta(\e)$ in the definition Eq.~\ref{eq:Cw} is replaced with the $2\pi$-periodic ``Lorentzian" with the width $\eta$: $\delta(\e) \rightarrow \sum_{k=-\infty}^{\infty} e^{ik\e-|k|\eta}/2\pi$~\cite{braun_self,Altland-rev,Haake}.
In the case where $\eta N \gtrsim 1$, we can focus on the dominant standard saddle point as its real part of the action at zero source field ($\re S^{(s)}=0$) is small compared with that of the nonstandard saddle point ($\re S^{(ns)}|_{J=0}=N\eta$)~\cite{Kamenev-GUE}.
In Ref.~\cite{entropy}, the smoothed two-point level correlation function $R_2^{(s)}(\w)$ was obtained from the consideration of the standard saddle point only. We will show that the inclusion of the nonstandard saddle point allows us to extract the fine structure of nearby quasienergy eigenstates.

For the standard saddle point, the quadratic fluctuations around it ($\delta Z=Z-Z^{(s)}=Z$) are governed by
\begin{align}\label{eq:dS2}
\begin{aligned}
\delta S_2^{(s)} [ \delta \TZ, \delta Z]
=\, &
\STr  \left( \delta \tilde{Z}\delta  Z \right) 
-
\STr  \left( \delta \tilde{Z} \alpha_J \mathbf{U} \delta Z \beta_J \mathbf{U}^{\dagger}\right) .
\end{aligned}
\end{align}
We approximate the corresponding contribution to the generating function by~\cite{entropy}
\begin{align}\label{eq:Zs}
	\mathcal{Z}^{(s)}[J]
	\approx
	\det(1+J^+)\det(1+J^-)
	\int \D(\TZ,Z) e^{-\left\langle \delta S_2^{(s)}\right\rangle }.
\end{align}
Here the ensemble averaged effective action $\left\langle \delta S_2^{(s)}\right\rangle$ can be further simplified to,
\begin{align}\label{eq:Seff2}
\begin{aligned}
	 \left\langle\delta S_2^{(s)} \right\rangle  
	=
	 \sum
	 s_{a}
	 \tilde{Z}_{ij}^{ab}  Z_{j'i'}^{ba} 
	 \left(
	 \delta_{ii'}\delta_{jj'}
	 -
	 (\alpha_J)^{bb}_{ji} (\beta_J)^{aa}_{i'j'}/N 
	 \right),
	 \end{aligned}
	\end{align}
where $s_{\msf{B/F}}=\pm 1$. We have used the fact that the second moment of the Floquet operator for the Floquet RQCs  under consideration is given by~\cite{entropy}\footnote{We note that the choice of the orthonormal basis $(\left\lbrace\ket{i}\right\rbrace)$ does not affect this result.}:
\begin{align}\label{eq:UU}
\left\langle U_{ij} U^{\dagger}_{j'i'} \right\rangle 
=
\frac{1}{N} \delta_{ii'}\delta_{jj'}.
\end{align}



Performing the Gaussian integration in Eq.~\ref{eq:Zs} and using Eq.~\ref{eq:C2}, we obtain
\begin{align}
\begin{aligned}\label{eq:Cstan-1}
\bar{C}_{nn'm'm}^{(s)}(\alpha_F,\beta_F)
=\,&
\delta_{nn'} \delta_{mm'} 
\left[ 
1+
\frac{1}{N^2} 
\left( 
\frac{\aF\bF}{1-\aF\bF}
\right)^2 
\right] 
\\
&
+
\delta_{nm} \delta_{n'm'}
\frac{1}{N} 
\frac{\aF\bF}{1-\aF\bF},
\end{aligned}
\end{align}
where we have set $\aB=\aF$ and $\bB=\bF$.
When substituted into Eq.~\ref{eq:C1}, this leads to the smoothed correlation function $C^{(s)}_{nn'm'm}(\w)$, 
\begin{subequations}
\begin{align}
&\begin{aligned}\label{eq:Cstan}
C_{nn'm'm}^{(s)}(\w)
=&
\delta_{nn'} \delta_{mm'} 
\left(
\frac{R_2^{(s)} (\w)}{N^2}
+
\frac{1}{2 \pi N^2} 
\right) 
-
\frac{\delta_{nm} \delta_{n'm'}}{2\pi N},
\end{aligned}
\\
&\begin{aligned}
R_2^{(s)} (\w)
=\,&
\frac{N^2}{2\pi}
-\frac{1}{4\pi}
\dfrac{1}{\sin^2 (\w/2)}.
\end{aligned}
\end{align}
\end{subequations}

The action for the quadratic fluctuations around the nonstandard saddle point ($\delta Z=Z-Z^{(ns)}$) can be reduced~\cite{Sup} to  an expression almost identical to that of the standard saddle point in Eq.~\ref{eq:dS2}, 
but with $\alpha_J^{\msf{FF}}$ and $\beta_J^{\msf{FF}}$ replaced by  $(\beta_J^{\msf{FF}})^{-1}=\bF^{-1} (1 +J^-)$ and $(\alpha_J^{\msf{FF}})^{-1}=\aF^{-1}(1+J^+)$, respectively.
As a result, the contribution from the nonstandard saddle point to $\bar{C}_{nn'm'm}$ can be deduced from that of the standard saddle point. 
It assumes the form
\begin{align}\label{eq:Cnonstan-1}
\begin{aligned}
&\bar{C}_{nn'm'm}^{(ns)}(\aF,\bF)
=
-\delta_{nn'} \delta_{mm'} 
\frac{1}{N^2}  
(\aF\bF)^N
\frac{\aF\bF}{(1-\aF \bF)^2}.
\end{aligned}
\end{align}

Combining the contributions of the standard and nonstandard saddle points (Eqs.~\ref{eq:Cstan-1} and~\ref{eq:Cnonstan-1}),  we find, with the help of Eq.~\ref{eq:C1}, the explicit expression for $C_{nn'm'm}(\w)$:
\begin{subequations}
 \begin{align}
&\begin{aligned}\label{eq:Cw-final}
&C_{nn'm'm}(\w)
=
\delta_{nn'}\delta_{mm'}
\left( 
\frac{R_2(\w)}{N^2} +\frac{1}{2\pi N^2}
\right) 
-
\frac{\delta_{nm}\delta_{n'm'}}{2 \pi N} 
\\
&
+
\left( 
\frac{1}{N}
\delta_{nm}\delta_{n'm'}
-
\frac{1}{N^2} 
\delta_{nn'} \delta_{mm'} 
\right) 
\delta(\w),
\end{aligned}
\\
&\begin{aligned}\label{eq:R2-cue}
&R_2(\w)
=
\frac{N^2}{2\pi}
-\frac{1}{2\pi}
\dfrac{\sin^2 (N \w/2)}{\sin^2 (\w/2)}
+
N \delta(\w).
\end{aligned}
\end{align}
\end{subequations}
We note that here $R_2(\w)$ is identical to the two-level correlation function for the CUE~\cite{Mehta,Haake}.
This result is equivalent to the random matrix prediction Eq.~\ref{eq:Cw-exact} to the leading order in $1/N$.
Moreover, the smoothed correlation function $C^{(s)}_{nn'm'm}(\w)$ in Eq.~\ref{eq:Cstan}, arising from the contribution of the standard saddle point only, is also consistent with Eq.~\ref{eq:Cw-exact} after replacing $R_2(\w)$ with its smoothed version $R_2^{(s)}(\w)$.
This is not surprising as the expression for the second moment of the Floquet operator (Eq.~\ref{eq:UU}) also holds for the CUE. 
At the quadratic level, the Floquet RQCs are therefore described by the same effective field theory as the CUE~\cite{entropy}.

In Eq.~\ref{eq:Cw-final} (and Eq.~\ref{eq:Cstan}), the second and third terms are of higher order in $1/N$ compared with the leading order first term there. 
They can be rewritten as $\delta_{nn'} \delta_{mm'} R_2^{(\msf{dis})}(\w)/N^4$ and $-\delta_{mn} \delta_{n'm'} R_2^{(\msf{dis})}(\w)/N^3$. Here $R_2^{(\msf{dis})}(\w)=N^2/2\pi$ denotes the disconnected part of the two-level correlation function, and is the dominating term in $R_2(\w)$ for $\w$ much larger than mean level spacing $2\pi/N$.
We note that they do not agree with the corresponding higher order terms in the RMT prediction Eq.~\ref{eq:Cw-exact}, i.e., $\delta_{nn'} \delta_{mm'} R_2(\w)/N^4$ and $-\delta_{mn} \delta_{n'm'} R_2(\w)/N^3$.
In fact, the higher order fluctuations ignored in the current calculation can give rise to contributions of the same orders as these terms for $\w \lesssim 2\pi/N$. 
In the Supplementary Material~\cite{Sup}, we investigate the role played by the higher order fluctuations of the supermatrix $Z$ in the sigma model for the CUE, as a simple example.
We recover the next leading higher order term $-\delta_{mn} \delta_{n'm'} R_2^{(s)}(\w)/N^3$ in the smoothed correlation function $C^{(s)}_{nn'm'm}(\w)$ for the CUE by inclusion of higher order fluctuations around the standard saddle point.

 Using our result for $C_{nn'm'm}(\w)$ in Eq.~\ref{eq:Cw-final}, one can immediately show that, for the current model, the SFF, the PSFF, and the two-point correlation functions of local traceless operators agree with the corresponding RMT predictions~\cite{Sup}.
In particular, we find that the SFF $K(t)$
of our current model
is given by 
the Fourier transform of $R_2(\w)$ in Eq.~\ref{eq:R2-cue}:
\begin{align}\label{eq:Kt-cue}
	K(t)=\sum_{n,m=1}^{N} C_{nnmm}(t)
	=\min(|t|,N) +N^2 \delta_{t,0}.
\end{align}
The PSFF~\cite{Zoller,Chalker-4} $K_{\mathcal{A}}(t)=\left\langle \Tr_{\bar{\mathcal{A}}} \left[ \Tr_\mathcal{A} U(t) \Tr_\mathcal{A} U^{\dagger}(t) \right] \right\rangle$
acquires a form
\begin{align}\label{eq:KA}
\begin{aligned}
K_{\mathcal{A}}(t) 
=&
\sum_{i_1^{\mathcal{A}},i_2^{\mathcal{A}}=1}^{N_{\mathcal{A}}}
\sum_{j_1^{\bar{\mathcal{A}}},j_2^{\bar{\mathcal{A}}}=1}^{N/N_{\mathcal{A}}}
C_{(i_1^{A}, j_1^{\bar{\mathcal{A}}}), (i_1^{A}, j_2^{\bar{\mathcal{A}}}),(i_2^{A}, j_2^{\bar{\mathcal{A}}}),(i_2^{A}, j_1^{\bar{\mathcal{A}}})}(t)
\\
\approx &
\frac{N_{\mathcal{A}}}{N} 
K(t)
+
\left( \frac{N}{N_{\mathcal{A}}}
-
\frac{N_{\mathcal{A}}}{N}\right) 
(1-\delta_{t,0}).
\end{aligned}
\end{align}
Here we use $\Tr_{\mathcal{A}}$ and $\Tr_{\bar{\mathcal{A}}}$ to indicate the trace operators over the Hilbert spaces of the subsystem $\mathcal{A}$ and its complement $\bar{\mathcal{A}}$, respectively.
$N_{\mathcal{A}} \gg 1$ denotes the dimension of the Hilbert space of the subsystem $\mathcal{A}$.
Two-component vector $(i^{A}, j^{\bar{\mathcal{A}}})$ labels the Hilbert space of the entire system, while its first and second components label the subsystem $\mathcal{A}$ and its complement $\bar{\mathcal{A}}$, respectively.
The second equality of Eq.~\ref{eq:KA} agrees with the PSFF of the CUE
of dimension $N_{\mathcal{A}} \gg 1$, which then indicates that the subsystem ETH~\cite{SUBETH} is satisfied in the current model~\cite{Zoller}.


Using Eq.~\ref{eq:Cw-final}, one also finds that the Fourier transform of the dynamical correlation functions of local traceless operators $O$ and $O'$, i.e., $C_{OO'}(t)=\frac{1}{N}\left\langle \Tr \left( O(t) O' \right) \right\rangle $, is given by
\begin{align}
\begin{aligned}
C_{OO'}(\w)
=
\!\!\!\!\!\!
\sum_{nn'm'm}
\!\!\!\!\!\!
\frac{O_{mn} O'_{n'm'}}{N}
C_{nn'm'm}(\w)
\approx
\frac{ R_2(\w)}{N^3} 
\Tr(OO'),
\end{aligned}
\end{align}
consistent with the RMT prediction.
We note that the contribution of higher order in $1/N$ terms in $C_{nn'm'm}(\w)$ to the operator correlation function $C_{OO'}(\w)$ may be non-negligible in some cases depending on the specific operators $O, O'$  and is responsible for the difference between our result and the RMT prediction for the case of traceful operators~\cite{Sup}.
As shown earlier, the fact that the autocorrelation function for a local operator $O$ (i.e., $C_{OO'}(\w)$ with $O=O'$) follows the RMT prediction is in agreement with ETH. 
It shows that $f(\w)$ for operator $O$ in the ETH ansatz matches the RMT prediction $f(\w) =\sqrt{\Tr(O^2)/N}$.

In summary, this letter investigates the statistics of quasienergy eigenstates for a family of Floquet random quantum circuits with Haar-distributed two-qudit random unitary gates. We compute the correlation function of energy eigenstates, i.e., $C_{nn'm'm}(\w)$ defined in Eq.~\ref{eq:Cw}, via a SUSY sigma-model calculation considering both the standard and nonstandard saddle points. We prove that $C_{nn'm'm}(\w)$ follows the RMT prediction Eq.~\ref{eq:Cw-exact} to the leading order in $1/N$, which shows that the SFF, the PSFF, and the correlation function of local traceless operators are in agreement with the corresponding RMT predictions. 
Moreover, this result is consistent with ETH and shows that $f(\w)$ entering the ETH ansatz for the matrix elements of operator $O$ also matches the RMT prediction $f(\w)=\sqrt{\Tr (O^2)/N}$.
Further analysis is required to investigate whether $R_{\mu\nu}$ in Eq.~\ref{eq:ETH}, for different pairs of eigenstates $\ket{\mu}, \ket{\nu}$,  follow independent Gaussian distribution, or are instead described by the general ETH formulation proposed in Ref.~\cite{Kurchan}.
See, for example, the numerical studies in Ref.~\cite{Chalker-ETH, Prosen-ETH}.
The current sigma model calculation can be generalized to compute the higher order correlation function of quasienergy eigenstates from the higher order derivatives of the generating function $\mathcal{Z}[J]$ with respect to the source $J$. 
These higher order correlation functions of quasienergy eigenstates are related to higher-order correlation functions of operators (such as the out-of-time-order correlators)
and contain extra information about the statistics of the matrix elements of operators required for a full analysis of the validity of the generalized ETH~\cite{Kurchan,ETH-Bound}.
We leave this for a future work.

This work was supported by the U.S. Department of Energy, Office of Science, Basic Energy
Sciences under Award No. DE-SC0001911. Y.L. acknowledges a post-doctoral fellowship
from the Simons Foundation “Ultra-Quantum Matter” Research Collaboration.

\bibliography{circuit}
	
\end{document}


\title{
		Field theory approach to eigenstate thermalization in random quantum circuits
		\\
		Supplemental Material
	}
	\author{Yunxiang Liao}
	\affiliation{Joint Quantum Institute, Department of Physics, University of Maryland, College Park, MD 20742, USA.}
	\affiliation{Condensed Matter Theory Center, Department of Physics, University of Maryland, College Park, MD 20742, USA.}
	\author{Victor Galitski}
	\affiliation{Joint Quantum Institute, Department of Physics, University of Maryland, College Park, MD 20742, USA.}
	
	\date{\today}

	\maketitle
	
	\tableofcontents
	
	
	\section{Introduction}
	
	In this Supplementary Material, we study the correlation function $C_{nn'm'm}(\w)$ defined as
	\begin{align}\label{eq:Cw}
	\begin{aligned}
	C_{nn'm'm}(\w)
	=
	\left\langle 
	\sum_{\nu,\mu}
	\left\langle n| \nu\right\rangle 
	\left\langle \nu |n'\right\rangle
	\left\langle m'|\mu\right\rangle
	\left\langle \mu|m \right\rangle
	\delta\left( \w- E_\nu+E_{\mu}\right) 
	\right\rangle .
	\end{aligned}
	\end{align}
	Here $\ket{\nu}$ represents the (quasi)energy eigenstate with (quasi)energy $E_{\nu}$, 
	and $\left\lbrace \ket{n}\right\rbrace $ denotes an arbitrary set of basis states of interest.
	The angular bracket indicates ensemble averaging.
	The Fourier transform of the correlation function $C_{nn'm'm}(\w)$  acquires the form
	\begin{align}\label{eq:Ct}
	\begin{aligned}
	&	C_{nn'm'm}(t)
	=
	\left\langle 
	U_{nn'}(t)
	U^{\dagger}_{m'm}(t)
	\right\rangle ,
	\end{aligned}
	\end{align}
	where $U_{nn'}(t)$ represents the matrix element of the time evolution operator at time $t$ (measured in units of period for Floquet systems) in the basis $\left\lbrace \ket{n} \right\rbrace $, i.e., $U_{nn'}(t)=\bra{n} U(t) \ket{n'}$.
	
    In Sec.~\ref{sec:RMT}, we provide  a brief derivation of the connection between the correlation function $C_{nn'm'm}(t)$ and the spectral form factor (SFF) $K(t)$ for the circular unitary ensemble (CUE). 
    In Sec.~\ref{sec:RQC}, we use a sigma-model approach to prove that, for a large family of Floquet random quantum circuits (where independently Haar distributed random unitary matrices are applied to pairs of neighboring qudits of local Hilbert space dimension $q\rightarrow \infty$), the correlation function $C_{nn'm'm}(\w)$ coincides with that of the CUE to the leading order in the Hilbert space dimension $N$.
    In Sec.~\ref{sec:HOF}, we investigate the higher order fluctuations of the sigma-model supermatrix field for the CUE, and compute their contribution to the CUE correlation function $C_{nn'm'm}(\w)$.
   Finally, in Sec.~\ref{sec:PP}, we use the obtained result for the correlation function $C_{nn'm'm}(t)$ to derive some of the physical properties of the Floquet random quantum circuits under consideration, including  the SFF, the partial spectral form factor (PSFF) and the dynamical correlation function of operators.

	\section{Correlation function of the quasienergy eigenstates for the CUE}\label{sec:RMT}
	
	Consider now an ensemble of Floque systems (with $U(t)=U^t$) whose Floquet operators $U$ are $N\times N$ random matrices drawn from the circular unitary ensemble (CUE).
	Diagonalizing the Floquet operator $U$ by a unitary matrix $V$:
	\begin{align}
	\begin{aligned}
		U=V 
        \msf{diag} \left\lbrace e^{-iE_1}, e^{-iE_2}, ..., e^{-iE_N}\right\rbrace 
        V^{\dagger},
	\end{aligned}
	\end{align}
	and using the fact that the quasienergy eigenvalues $\left( \left\lbrace E_{i} \right\rbrace\right) $ and eigenstates (columns of the diagonalizing matrix $V$) of the CUE are statistically independent, one can rewrite $C_{nn'm'm}(\w)$ as
	\begin{align}\label{eq:C-CUE-1}
	\begin{aligned}
		C_{nn'm'm}(\w)
		=
		\sum_{\nu,\mu=1}^{N}
		\int_{\mathcal{U}(N)} dV
		V_{n\nu} V^{\dagger}_{\nu n'}
		V_{m'\mu} V^{\dagger}_{\mu m}
		\int \prod_{\gamma=1}^{N} dE_{\gamma}
		P(E_1,E_2,...,E_N)
		\delta\left( \w- E_\nu+E_{\mu}\right) .
	\end{aligned}
	\end{align}
	Here $dV$ denotes the Haar measure on the unitary group $\mathcal{U}(N)$ and is normalized by $\int_{\mathcal{U}(N)} dV=1$. $P(E_1,E_2,...,E_N)$ represents the joint probability distribution of the quasienergies  $\left\lbrace E_{\gamma} \right\rbrace$.
	
	The Haar integral in the equation above can be calculated with the help of the Weingarten calculus~\cite{Samuel, Weingarten, collins2003,collins2006,collins2021,Weingarten-rev,Zee, Beenakker}:
	\begin{align}\label{eq:Weingarten}
	\begin{aligned}
	&\int_{\mathcal{U}(N)} d V
	V_{i_1 j_1} V_{i_2 j_2} V^{\dagger}_{j_1' i_1'} V^{\dagger}_{j_2' i_2'}
	\\
	=&
	\frac{1}{N^2-1} 
	\left( 
	\delta_{ i_1, i_1'} \delta_{i_2, i_2'} \delta_{ j_1, j_1'} \delta_{j_2, j_2'} 
	+
	\delta_{ i_1, i_2'} \delta_{i_2, i_1'} \delta_{ j_1, j_2'} \delta_{j_2, j_1'} 
	\right) 
	-
	\frac{1}{N(N^2-1)} 
	\left( 
	\delta_{ i_1, i_1'} \delta_{i_2, i_2'} \delta_{ j_1, j_2'} \delta_{j_2, j_1'} 
	+
	\delta_{ i_1, i_2'} \delta_{i_2, i_1'} \delta_{ j_1, j_1'} \delta_{j_2, j_2'} 
	\right).
	\end{aligned}
	\end{align}
	Inserting Eq.~\ref{eq:Weingarten} into Eq.~\ref{eq:C-CUE-1}, one obtains
	\begin{align}\label{eq:Cw-exact}
	\begin{aligned}
	C_{nn'm'm} (\w)
	=
	&
	\frac{1}{N^2-1} 
	R_2(\w)
	\left(
	\delta_{nn'}\delta_{mm'}
	-
	\frac{1}{N} \delta_{mn} \delta_{n'm'}
	\right)
	+
	\left( 
	\frac{N}{(N^2-1)} 
	\delta_{mn} \delta_{n'm'}
	-
	\frac{1}{(N^2-1)} 
	\delta_{nn'}\delta_{mm'}
	\right) \delta(\w),
	\end{aligned}
	\end{align}
	where $R_2(\w)$ denotes the two-level correlation function defined as
	\begin{align}\label{eq:R2}
	R_2(\w)
	=
	\left\langle 
	\sum_{\nu,\mu=1}^{N}
	\delta\left( \w- E_\nu+E_{\mu}\right) 
	\right\rangle.
	\end{align}
	For the CUE, it is known that $R_2(\w)$ is given by~\cite{Mehta,Haake}:
	\begin{align}\label{eq:R2-cue}
	R_2(\w)
	=
	\frac{N^2}{2\pi}
	-\frac{1}{2\pi}
	\dfrac{\sin^2 (N \w/2)}{\sin^2 (\w/2)}
	+
	N \delta(\w).
	\end{align}
	
	In the Fourier representation, Eq.~\ref{eq:Cw-exact} becomes
	\begin{align}\label{eq:Ct-exact}
	\begin{aligned}
	C_{nn'm'm} (t)
	=
	&
	\frac{1}{N^2-1} 
	K(t)
	\left(
	\delta_{nn'}\delta_{mm'}
	-
	\frac{1}{N} \delta_{mn} \delta_{n'm'}
	\right)
	+
	\frac{N}{(N^2-1)} 
	\delta_{mn} \delta_{n'm'}
	-
	\frac{1}{(N^2-1)} 
	\delta_{nn'}\delta_{mm'},
	\end{aligned}
	\end{align}
	where $K(t)$ indicates the spectral form factor (SFF):
	\begin{align}\label{eq:Kt}
		K(t)
		=
		\left\langle 
		\sum_{\nu,\mu=1}^{N}
		e^{-iE_{\mu}t+iE_{\nu}t}
		\right\rangle
		=
		\left\langle 
		\Tr U(t)
		\Tr U^{\dagger}(t)
		\right\rangle.
	\end{align}
	For the CUE ensemble, $K(t)$ acquires the form
	\begin{align}\label{eq:Kt-cue}
		K(t)=\min(|t|,N)+N^2\delta_{t,0}.
	\end{align}

    In an analogous manner, we now investigate the statistics of the matrix elements of an arbitrary Hermitian operator $O$ in the basis of the quasienergy eigenstates of the CUE, i.e., $O_{\nu\mu}\equiv \bra{\nu}O\ket{\mu}$. The ensemble average of $O_{\nu\mu}$ is given by
    \begin{align}\label{eq:O1}
    \begin{aligned}
        \left\langle
        O_{\nu\mu}
        \right\rangle
        =
        \sum_{n,m=1}^{N}
        O_{nm}
		\int_{\mathcal{U}(N)} dV
		V^{\dagger}_{\nu n} V_{m\mu}    
        =
        \frac{1}{N}\Tr (O) \delta_{\nu\mu}.
    \end{aligned}
    \end{align}
    Here we have used the following Haar integral
    \begin{align}\label{eq:Weingarten}
    \begin{aligned}
	&\int_{\mathcal{U}(N)} d V\,
	V_{i j} V^{\dagger}_{j' i'} 
	=
	\frac{1}{N} 
	\delta_{ i, i'}  \delta_{j, j'}.
	\end{aligned}
	\end{align}
    Similarly, we find that the variance of the matrix element $O_{\nu\mu}$ is
    \begin{align}\label{eq:O2}
    \begin{aligned}
        \left\langle
        |O_{\nu\mu}|^2
        \right\rangle
        =&
        \sum_{n,m,n',m'=1}^{N}
        O_{nm} O_{m'n'}
		\int_{\mathcal{U}(N)} dV
		V^{\dagger}_{\nu n} V_{m\mu}
        V^{\dagger}_{\mu m'} V_{n'\nu}
        \\
        =&
       \left[
        \frac{1}{N^2-1}\Tr(O^2) 
        -
        \frac{1}{N(N^2-1)}\Tr^2(O)
        \right]
        +
        \delta_{\nu\mu}
        \left[
        \frac{1}{N^2-1}\Tr^2(O)
        -
        \frac{1}{N(N^2-1)}\Tr(O^2)
        \right].
    \end{aligned}
    \end{align}
    Combining Eqs.~\ref{eq:O1} and~\ref{eq:O2}, one can express $O_{\nu\mu}$ as
    \begin{align}\label{eq:O3}
    \begin{aligned}
        |O_{\nu\mu}|
        = &
        \frac{1}{N}\Tr (O) \delta_{\nu\mu}
        +
        \left\lbrace
       \left[
        \frac{1}{N^2-1}\Tr(O^2) 
        -
        \frac{1}{N(N^2-1)}\Tr^2(O)
        \right]
        \right.
        \\
        &
        +
        \left.
        \delta_{\nu\mu}
        \left[
        \left(\frac{1}{N^2-1}-\frac{1}{N^2}\right)\Tr^2(O)
        -
        \frac{1}{N(N^2-1)}\Tr(O^2)
        \right]
        \right\rbrace^{1/2}
        R_{\nu\mu}
        \\
        \approx &
        \frac{1}{N}\Tr (O) \delta_{\nu\mu}
        +
       \left[
        \frac{1}{N^2}\Tr(O^2) 
        -
        \frac{1}{N^3}\Tr^2(O)
        \right]^{1/2}
        R_{\nu\mu},
    \end{aligned}
    \end{align}
    where in the last equality we kept only the leading order terms in $1/N$.
    
	From the derivations above, it is straightforward to see that Eqs.~\ref{eq:Cw-exact},~\ref{eq:Ct-exact} and~\ref{eq:O3} are also applicable to an ensemble of Hamiltonian systems whose time-evolution operators $U(t)=e^{-iHt}$ are generated by Hermitian Hamiltonians $H$ drawn from the Gaussian unitary ensemble (GUE), see also Ref.~\cite{Cotler2020,Cotler2017,ETH-review}.

	
	\section{Correlation function of the quasienergy eigenstates for the Floquet random quantum circuits}\label{sec:RQC}
	
	In this section, we first construct a supersymmetric (SUSY) sigma model for the calculation of the correlation function $C_{nn'm'm}(\w)$, applicable to any ensemble of Floquet systems. We then employ this sigma model to a large family of Floquet random quantum circuits which consist of qudits in an arbitrary dimensional cubic lattice with either periodic or open boundary condition. Within each period, random unitary matrices drawn independent from the CUE are applied to different pairs of neighboring qudits at various substeps. 
    We find the correlation function $C_{nn'm'm}(\w)$ for any model in this family agrees with the RMT prediction Eq.~\ref{eq:Cw-exact} in the limit of large single-particle Hilbert space dimension $q\rightarrow \infty$, irrespective of the ordering of the local gates.
   This sigma model calculation is a generalization to that of the  two-level correlation function $R_2(\w)$ in Ref.~\cite{entropy}, see also the review~\cite{Altland-rev} and the references therein.
	
	\subsection{Preliminaries}
	
	Consider first an arbitrary ensemble of Floquet systems.
	To evaluate the correlation function $C_{nn'm'm}(\w)$, one can make use of the generating function
	\begin{align}
	\begin{aligned}
	\mathcal{Z}[J]
	=&
	\left\langle 
	\int_0^{2\pi} \frac{d\phi}{2\pi}
	\dfrac
	{\det \left( 1-\aF e^{i\phi} U+ J^+\right) 
		\det \left( 1-\bF e^{-i\phi} U^{\dagger} +J^-\right) 
	}
	{\det \left( 1-\aB e^{i\phi} U\right) 
		\det \left( 1-\bB e^{-i\phi} U^{\dagger}\right) 
	}
	\right\rangle.
	\end{aligned}
	\end{align}
	Here the angular bracket represents averaging over the ensemble of the Floquet operators $U$.
	$J^{\pm}$ is a $N\times N$ source matrix, with $N$ being the dimension of Hilbert space wherein the Floquet operator acts.
	
	Taking derivatives of the generating function $\mathcal{Z}[J]$ with respect to the matrix elements of the source field $J^{\pm}$, and then setting $J=0$, $\aB=\aF$ and $\bB=\bF$
    \footnote{In principal, one can directly set $\aB=\aF$ and $\bB=\bF$ at the beginning of the calculation. However, as will become apparent later, it is more convenient to apply this procedure at the end.},
	we have
	\begin{align}\label{eq:C1}
	\begin{aligned}
	\bar{C}_{nn'm'm}(\aF, \bF)
	\equiv\, &
	\dfrac{\partial^2\mathcal{Z}[J]}{\partial J^+_{n'n}\partial J^-_{mm'}}
	\bigg\lvert_{J=0,\aB=\aF,\bB=\bF}
	=
	\sum_{k=0}^{\infty}
	C_{nn'm'm}(k)
	\left( \aF\bF\right)^k,
	\end{aligned}
	\end{align}
	where $C_{nn'm'm}(k)$ is defined in  Eq.~\ref{eq:Ct} with $k$ being the discrete time measured in units of period.
	We can therefore see that its Fourier transform $C_{nn'm'm}(\w)$ (Eq.~\ref{eq:Cw}) can be obtained from $\bar{C}_{nn'm'm}(\aF, \bF)$
	when the product of $\aF$ and $\bF$ is set to $\aF\bF=e^{i\w}$:
	\begin{align}\label{eq:C2}
	\begin{aligned}
		C_{nn'm'm}(\w)
		=
		\frac{1}{2\pi}
		\left( 
		\bar{C}_{nn'm'm}(\aF, \bF)
		+
		\bar{C}^{*}_{n'nmm'}(\aF, \bF)
		-
		\delta_{nn'}\delta_{mm'}
		\right) .
	\end{aligned}
	\end{align}
	
	\subsection{Sigma model}
	
	The starting point of the sigma model derivation is the following Gaussian superintegral representation of the generating function:
	\begin{align}\label{eq:ZJ}
	\begin{aligned}
	&\mathcal{Z}[J]
	=
	\left\langle 
	\int_0^{2\pi} \frac{d\phi}{2\pi}
	\int \D (\bpsi,\psi)
	\exp
	\left\lbrace 
	\begin{aligned}
	&-\bar{\psi}^{+,F}_{i} \left( \delta_{ij}-\aF e^{i\phi}U_{ij} +J^{+}_{ij}\right)  \psi^{+,F}_{j}
	-\bar{\psi}^{-,F}_{i} \left( \delta_{ij}-\bF e^{-i\phi}U^{\dagger}_{ij} +J^{-}_{ij}\right)  \psi^{-,F}_{j}
	\\
	&-\bar{\psi}^{+,B}_{i} \left( \delta_{ij}-\aB e^{i\phi}U_{ij} \right)  \psi^{+,B}_{j}
	-\bar{\psi}^{-,B}_{i} \left( \delta_{ij}-\bB e^{-i\phi}U^{\dagger}_{ij} \right)  \psi^{-,B}_{j}
	\end{aligned}
	\right\rbrace 
	\right\rangle .
	\end{aligned}
	\end{align}
	Here $\psi^{s,a}_{i}$ carries an index $s=\pm$ labeling the retarded(+)/advanced(-) space associated with the forward/backward time evolution $U/U^{\dagger}$, an index $a=\msf{B},\msf{F}$ labeling the Boson($\msf{B}$)/Fermion($\msf{F}$) space, and $i=1,2,...,N$ which labels the Hilbert ($\mathcal{H}$) space. The integration measure $\D(\bpsi,\psi)$ is normalized such that the generating function $\mathcal{Z}=1$ when $J=0$ and $\aF=\aB=\bF=\bB=0$.
	
	For simplicity, we introduce $2N\times 2N$ matrices $\alpha_J$ and $\beta_J$ in the direct product of Boson-Fermion and Hilbert spaces ($\msf{BF} \otimes \mathcal{H}$):
	 \begin{align}
	\begin{aligned}
	\alpha_J
	=
	\begin{bmatrix}
	\aB \mathbb{1}_{\mathcal{H}} & 0
	\\
	0 & \aF(1+J^+)^{-1}
	\end{bmatrix}_{\msf{BF}},
	\qquad
	\beta_J
	=
	\begin{bmatrix}
	\bB \mathbb{1}_{\mathcal{H}}  & 0
	\\
	0 & \bF(1+J^-)^{-1}
	\end{bmatrix}_{\msf{BF}},
	\end{aligned}
	\end{align}
	and rewrite
	 Eq.~\ref{eq:ZJ} as
	\begin{align}
	\begin{aligned}
	\mathcal{Z}[J]
	=&
	\det(1+J^+) 
	\det(1+J^-)
	\\
	&\times
	\left\langle 
	\int_0^{2\pi} \frac{d\phi}{2\pi}
	\int
	\D (\bar{\psi},\psi)
	\exp
	\left\lbrace 
	\begin{aligned}
	-\begin{bmatrix}
	\bar{\psi}^{+}
	&
	\bar{\psi}^{-}
	\end{bmatrix}
	\begin{bmatrix}
	1-e^{i\phi}\alpha_J \mathbf{U} & 0
	\\
	0 & 1-e^{-i\phi}\beta_J \mathbf{U}^{\dagger}
	\end{bmatrix}_{\msf{RA}}
	\begin{bmatrix}
	{\psi}^{+}
	\\
	{\psi}^{-}
	\end{bmatrix}
	\end{aligned}
	\right\rbrace 
	\right\rangle .
	\end{aligned}
	\end{align}
	Here the subscript in $[.]_{\msf{BF}}$ ($[.]_{\msf{RA}}$) indicates that the matrix is expressed in the Boson-Fermion (retarded-advanced) space.
   We use $\mathbb{1}_{\mathcal{H}}$ ($\mathbb{1}_{\msf{BF}}$) to denote the identity matrix in the Hilbert (Boson-Fermion) space.
   For simplicity, we also define $\mathbf{U}=U\otimes \mathbb{1}_{\msf{BF}}$ .
	
	The next step is applying the color-flavor transformation~\cite{Zirnbauer_1996,Zirnbauer_1998,Zirnbauer_1999,Zirnbauer_2021}  to convert the integration over the center phase $\phi$ to an integration over $2N\times2N$ supermatrices $Z$ and $\tilde{Z}$:
	\begin{align}\label{eq:ZJ-1}
	\begin{aligned}
	\mathcal{Z}[J]
	=&
	\det(1+J^+) 
	\det(1+J^-)
	\\
	& \times
	\left\langle 
	\int D(\tilde{Z},Z)
	\exp \left[ \STr \ln (1-\tilde{Z} {Z}) \right] 
	\int
	\D (\bpsi,\psi)
	\exp
	\left\lbrace 
	\begin{aligned}
	-\begin{bmatrix}
	\bar{\psi}^{+}
	&
	\bar{\psi}^{-}
	\end{bmatrix}
	\begin{bmatrix}
	1 & Z\beta_J \mathbf{U}^{\dagger}
	\\
	\tilde{Z}\alpha_J \mathbf{U}  & 1
	\end{bmatrix}_{\msf{RA}}
	\begin{bmatrix}
	{\psi}^{+}
	\\
	{\psi}^{-}
	\end{bmatrix}
	\end{aligned}
	\right\rbrace 
	\right\rangle.
	\end{aligned}
	\end{align}
	Here $Z^{ab}_{ij}$ carries indices $a,b=\msf{B},\msf{F}$ that label the Boson-Fermion space and indices $i,j=1,2,..,N$ that label the Hilbert space. The integration runs over all $2N\times2N$ complex supermatrices $Z$ and $\tilde{Z}$ with the constraints: $\tilde{Z}^{\msf{FF}}=-Z^{\msf{FF}}\,^{\dagger}$, $\tilde{Z}^{\msf{BB}}=Z^{\msf{BB}}\,^{\dagger}$, and $|Z^{\msf{BB}}\,^{\dagger}Z^{\msf{BB}}|<1$.
	$\STr$ represents the supertrace over the $\msf{BF}\otimes \mathcal{H}$ space.
	The integration measure $\D(\tilde{Z}, Z)$ is also normalized by $\mathcal{Z}[J=0]=1$ when $\aF=\aB=\bF=\bF=0$.

	Carrying out the integration over $\psi$ and $\bar{\psi}$, we arrive at the SUSY sigma model representation of the generating function:
	\begin{align}\label{eq:sigma}
	\begin{aligned}
		\mathcal{Z}[J]
		=&
		\det(1+J^+)\det(1+J^-)
		\left\langle 
		\int D(\tilde{Z},Z)
		\exp(-S[\TZ,Z])
		\right\rangle ,
		\\
		S[\TZ, Z]
		=&
		-
		\STr \ln (1-\tilde{Z} {Z})
		+
		\STr \ln \left[ 1- \tilde{Z}\alpha_J \mathbf{U} Z\beta_J \mathbf{U}^{\dagger} \right].
	\end{aligned}
	\end{align}
	We emphasize that this sigma model is exact and is applicable to an arbitrary ensemble of Floquet systems. 
	
	To compare with the sigma model for disordered Hamiltonian systems~\cite{Efetov}, one can make the transformation
	\begin{align}\label{eq:Q}
	Q=
	T\Lambda T^{-1},
	\qquad
	T=
	\begin{bmatrix}
	1  & Z
	\\
	\TZ  & 1
	\end{bmatrix}_{\msf{RA}},
	\qquad
	\Lambda
	=
	\begin{bmatrix}
	1  & 0
	\\
	0 & -1
	\end{bmatrix}_{\msf{RA}},
	\end{align}
	after which the action becomes
	\begin{align}\label{eq:SQ}
	\begin{aligned}
	&S[Q]
	=
	\STr\ln 
	\left[ 
	\left( 
	1-
	\begin{bmatrix}
	\alpha_J \mathbf{U} & 0
	\\
	0 & \beta_J \mathbf{U}^{\dagger}
	\end{bmatrix}
	\right)
	Q\Lambda
	+
	\left( 
	1+
	\begin{bmatrix}
	\alpha_J \mathbf{U} & 0
	\\
	0 & \beta_J \mathbf{U}^{\dagger}
	\end{bmatrix}
	\right)
	\right] .
	\end{aligned}
	\end{align}
	Supermatrix field $Q$ stays on the manifold with the constraints $\STr Q=0$ and $Q^2=1$.
	
	\subsection{Saddle points}
	
	Starting from the sigma model Eq.~\ref{eq:sigma}, we now apply it to the Floquet random quantum circuits under study.
	We consider the large $N\rightarrow \infty$ limit, and focus on the saddle points and quadratic fluctuations around them.  In Ref.~\cite{entropy}, we have shown that the contribution from higher-order fluctuations for the current model is of higher order in $1/N$.

	Taking variation of the action $S[\TZ,Z]$  (Eq.~\ref{eq:sigma}) with respect to $\TZ$ yields the saddle point equation for an arbitrary configuration of Floquet operator $U$:
	\begin{align}\label{eq:SPEQ}
	\begin{aligned}
	Z(1-\tilde{Z} {Z})^{-1}
	=
	\alpha_J \mathbf{U} Z\beta_J  
	\mathbf{U}^{\dagger} 
	\left( 1- \tilde{Z} \alpha_J \mathbf{U}  Z \beta_J  \mathbf{U}^{\dagger}\right)^{-1}.
	\end{aligned}
	\end{align}
	Note that, in Eq.~\ref{eq:sigma}, an averaging over the ensemble of  Floquet operator $U$ is required. We only look for solutions that solve the saddle point equation for all Floquet operators in the ensemble under consideration, and ignore all $U$ dependent solution assuming they are not important to the ensemble averaged field theory. 
		
	It is easy to see that 
	\begin{align}\label{eq:stan}
	\begin{aligned}
		Z^{(s)}=\TZ^{(s)}=0,
	\end{aligned}
	\end{align}
	is a solution to the saddle point equation.
	The corresponding $Q^{(s)}$ matrix which is related to $Z^{(s)},\TZ^{(s)}$ by Eq.~\ref{eq:Q} assumes the form
	\begin{align}
	\begin{aligned}
	Q^{(s)}=\sigma^3_{\msf{RA}} \otimes \mathbb{1}_{\msf{BF}} \otimes \mathbb{1}_{\mathcal{H}},
	\end{aligned}
	\end{align}
	where $\sigma^3_{\msf{RA}}$ is the Pauli matrix acting in  the retarded-advanced space.

	Another solution to the saddle point equation takes the form
	\begin{align}\label{eq:AA}
	\begin{aligned}
	Z^{(ns)}
	=
	\begin{bmatrix}
	0 & 0
	\\
	0 & z 
	\end{bmatrix}_{\msf{BF}},
	\qquad
	\TZ^{(ns)}
	=
	\begin{bmatrix}
	0 & 0
	\\
	0 & -z^{\dagger}
	\end{bmatrix}_{\msf{BF}},
	\end{aligned}
	\end{align}
	where $z$ is a $N \times N$ matrix that solves the following equation
		\begin{align}\label{eq:SPEQz}
		\begin{aligned}
		z(1+z^{\dagger}z)^{-1}
		=
		\aF\bF(1+J^+)^{-1} U  z (1+J^-)^{-1}  U^{\dagger} (1+\aF \bF z^{\dagger} (1+J^+)^{-1} U  z (1+J^-)^{-1}  U^{\dagger} )^{-1}.
		\end{aligned}
		\end{align}
The solution to the equation above can be expressed as $z = c_0 z'$ where $c_0'\rightarrow \infty$ and $z'$ is an arbitrary invertible matrix of order unity.
	The corresponding $Q^{(ns)}$ matrix for the nonstandard saddle point is given by
	\begin{align}\label{eq:QAA}
	\begin{aligned}
	Q^{(ns)}
	=
	\sigma^3_{\msf{RA}} \otimes \sigma^3_{\msf{BF}} \otimes \mathbb{1}_{\mathcal{H}},
	\end{aligned}
	\end{align}
	where $\sigma^3_{\msf{BF}}$ represents the Pauli matrix in the Fermion-Boson space.
	We note that $Q^{(ns)}$ is similar to the non-trivial saddle point considered by Andreev and Altshuler~\cite{Andreev-Altshuler} in the calculation of the two-level correlation function $R_2(\w)$ for disordered Hamiltonian systems.
	 
	In the following, we call $Z^{(s)}=0$ the standard saddle point and the non-trivial solution $Z^{(ns)}$ the nonstandard saddle point.
	Inserting Eq.~\ref{eq:stan} and Eq.~\ref{eq:AA} into Eq.~\ref{eq:sigma}, and making use of the saddle point equation, we find the actions at these saddle points 
	\begin{subequations}\label{eq:Ssaddle}
	\begin{align}
		&\begin{aligned}\label{eq:Ssaddle-s}
		S^{(s)}\equiv
    S[\TZ^{(s)}, Z^{(s)}]
		= \, & 
		0,
		\end{aligned}
		\\
		&\begin{aligned}\label{eq:Ssaddle-ns}
			S^{(ns)}\equiv 
    S[\TZ^{(ns)}, Z^{(ns)}]
			=\, & 
			-\Tr \ln \left[  \aF\bF (1+J^+)^{-1}  (1+J^-)^{-1} \right],
		\end{aligned}
	\end{align}
	\end{subequations}
	where $\Tr$ denotes the trace over the Hilbert space only. Note that the Floquet operator $U$ disappears  from the saddle point actions.
	
	When the source field $J^{\pm}$ is set to zero, the action at the nonstandard saddle point assumes the value of
	$
	S^{(ns)}|_{J=0}= -N \ln \left(  \aF \bF\right) 
	$.
	If we introduce an infinitesimal imaginary part $\eta$ to $\ln (\aF\bF)=i\w-\eta$ with $\eta N\gtrsim 1$, the real part of the action at the nonstandard saddle point $\re S^{(ns)}=N\eta$ is larger compared with the one at the standard saddle point $\re S^{(s)}=0$. 
	In this case, it is enough to take into account the dominant saddle point, i.e., the standard saddle point, as well as the quadratic fluctuations around it. The correlation function $C_{nn'm'm}(\w)$ obtained from derivatives of the generating function $\mathcal{Z}[J]$ is now
	smoothed over.
	More specifically, from Eqs.~\ref{eq:C1} and~\ref{eq:C2}, one can see that introducing an infinitesimal imaginary part $\eta$ to $\ln (\aF\bF)=i\w-\eta$ replace the $2\pi$-periodic delta function  $\delta(\w)$ in the definition of $C_{nn'm'm}(\w)$ (Eq.~\ref{eq:Cw}) with a $2\pi$-periodic  Lorentzian of width $\eta$, i.e., $\delta(\w)=\sum_{n=-\infty}^{\infty} e^{in\w}/2\pi \rightarrow \sum_{n=-\infty}^{\infty} e^{in\w-|n|\eta}/2\pi$.
	$C_{nn'm'm}(\w)$ now measures the correlation of pairs of quasienergy eigenstates with quasienergy separations $E_{\nu}-E_{\mu} $ around $\w$ weighted by the Lorentzian.
	Therefore, we can see that consideration of only standard saddle point yields the smoothed correlation function $C_{nn'm'm}(\w)$.
	
	\subsection{Quadratic fluctuations around the standard saddle point}
	
	The quadratic fluctuations around the standard saddle point (Eq.~\ref{eq:stan}) are governed by the action
		\begin{align}\label{eq:S2stan-0}
		\begin{aligned}
		\delta S_2^{(s)} [ \TZ, Z]
		=\, &
		\STr  \left[\tilde{Z} Z \right] 
		-
		\STr  \left[  \tilde{Z} \alpha_J \mathbf{U}  Z \beta_J \mathbf{U}^{\dagger}\right].
		\end{aligned}
		\end{align}
	Using Eq.~\ref{eq:Ssaddle-s} and ignoring the higher order fluctuations whose contributions are of higher order in $1/N$~\cite{entropy}, we approximate the contribution from the standard saddle point to the generating function $\mathcal{Z}[J]$ by
	\begin{align}\label{eq:Seff2}
	\begin{aligned}
	&\mathcal{Z}^{(s)}[J]
	\approx 
	\det(1+J^+) \det(1+J^-)
	\int \D(\TZ,Z) e^{-S_{\msf{eff}}^{(s)} [ \TZ, Z]},
	\\
	&S_{\msf{eff}}^{(s)} [ \TZ, Z]
	= 
	 \left\langle\delta S_2^{(s)}[\TZ, Z ] \right\rangle  
	=
	 \sum_{i,j,i',j'} \sum_{a,b=\msf{B},\msf{F}}
	 s_{a}
	 \tilde{Z}_{ij}^{ab}  Z_{j'i'}^{ba} 
	 \left(
	 \delta_{ii'}\delta_{jj'}
	 -
	 \frac{1}{N} (\alpha_J)^{bb}_{ji} (\beta_J)^{aa}_{i'j'}
	 \right),
	 \end{aligned}
	\end{align}
	where $s_{\msf{B}/\msf{F}}=\pm 1$.
	We have used  the fact that,
	for the Floquet random quantum circuits under consideration~\cite{entropy},
		\begin{align}\label{eq:U2}
		\left\langle U_{ij} U^{\dagger}_{j'i'} \right\rangle 
		=
		\frac{1}{N} \delta_{ii'}\delta_{jj'},
		\end{align}
	in an arbitrary orthonormal basis.
	Note that this equation also holds for the CUE of dimension $N$. Therefore, the quadratic fluctuations around the standard saddle point for the current model is described by the same ensemble averaged effective field theory as that of the CUE.
	
	To proceed, we divide $Z$ into two parts which involve, respectively, its diagonal and off-diagonal components in the Hilbert space:
	\begin{align}
	\begin{aligned}
	&Z=X+Y,
	\qquad
	X^{ab}_{ij}=\delta_{ij} Z_{ii}^{ab},
	\qquad
	Y^{ab}_{ij}= (1- \delta_{ij}) Z_{ij}^{ab},
	\end{aligned}
	\end{align}
	and similarly for $\TZ$.
	We then perform a Fourier transformation of the diagonal components $X$ and $\tilde{X}$:
	\begin{align}
	\begin{aligned}
	&X^{ab}(k)
	=
	\sum_{j=1}^{N}
	X^{ab}_{jj} e^{-i2\pi k j /N},
	\qquad
	\tilde{X}^{ab}(k)
	=
	\sum_{j=1}^{N}
	\tilde{X}^{ab}_{jj} e^{i2\pi k j/N}.
	\end{aligned}
	\end{align}
	
	In terms of $X(k)$, $\tilde{X}(k)$, $Y_{ij}$ and $\tilde{Y}_{ij}$, the ensemble averaged effective action $S_{\msf{eff}}^{(s)}$ in Eq.~\ref{eq:Seff2} can be expressed as
	\begin{align}\label{eq:S2stan}
	\begin{aligned}
	S_{\msf{eff}}^{(s)} [X,Y,\tilde{X},\tilde{Y}]
	=	&
	\sum_{a,b=\msf{B},\msf{F}}
	\sum_{i\neq j} 
	s_{a} \tilde{Y}_{ij}^{ab}  Y_{ji}^{ba} 
	+
	\frac{1}{N}
	\sum_{a,b=\msf{B},\msf{F}}
	\sum_{k=0}^{N-1} s_{a} \tilde{X}^{ab} (k) X^{ba} (k) 
	\\
	&-\frac{1}{N} 
	\sum_{a,b=\msf{B},\msf{F}}
	s_{a}\alpha_b 	\beta_a 
	\left( 
	\tilde{X}^{ab} (0) 
	-
	\delta_{bF} \frac{1}{N}
	\sum_{k=0}^{N-1}\tilde{X}^{aF} (k)  \sum_{i} e^{-i2\pi k i/N} J^+_{ii} 
	-
	\delta_{bF} 
	\sum_{i\neq j} \tilde{Y}_{ij}^{aF} J^+_{ji}  
	\right)
	\\
	&
	\times 
	\left( 
	X^{ba} (0) 
	-
	\delta_{aF} \frac{1}{N}
	\sum_{k=0}^{N-1} X^{bF} (k)  \sum_{i'} e^{i2\pi k i'/N} J^-_{i'i'} 
	-
	\delta_{aF} 
	\sum_{i'\neq j'} Y_{j'i'}^{bF} J^-_{i'j'} 
	\right) .
	\end{aligned}
	\end{align}
	Here we have employed the approximation $(1+J^{\pm})^{-1}\approx 1-J^{\pm}$ for small source matrix field $J$.
	
	When the source field is set to $J=0$, the action $S_{\msf{eff}}^{(s)}$ reduces to
	\begin{align}\label{eq:S2stan20}
	\begin{aligned}
	S_{\msf{eff}}^{(s)} [X,Y,\tilde{X},\tilde{Y}]|_{J=0}
	=	&
	\sum_{a,b=\msf{B},\msf{F}}
	\sum_{i\neq j} 
	s_{a} \tilde{Y}_{ij}^{ab}  Y_{ji}^{ba} 
	+
		\frac{1}{N}
	\sum_{a,b=\msf{B},\msf{F}}
	\sum_{k=0}^{N-1} s_{a} \tilde{X}^{ab} (k) X^{ba} (k) 
	(1-\delta_{k,0}\alpha_b \beta_a  ).
	\end{aligned}
	\end{align}
	The bare propagators for $X$ and $Y$ arising from this action are given by
	\begin{align}\label{eq:bareProp}
	\begin{aligned}
	&
	\left\langle X^{b'a'} (k')\tilde{X}^{ab} (k) \right\rangle_0
	=
	s_as_b \left\langle \tilde{X}^{ab} (k) X^{b'a'} (k') \right\rangle_0
	=
	s_a \dfrac{N}{1-\alpha_b\beta_a \delta_{k,0} }
	\delta_{kk'}\delta_{aa'}\delta_{bb'},
	\\
	&
	\left\langle Y_{j'i'}^{b'a'}\tilde{Y}_{ij}^{ab}  \right\rangle_0
	=
	s_a s_b\left\langle \tilde{Y}_{ij}^{ab} Y_{j'i'}^{b'a'}   \right\rangle_0
	=
	s_a
	\delta_{ii'}\delta_{jj'}\delta_{aa'}\delta_{bb'}.
	\end{aligned}
	\end{align}
	Here the angular bracket with subscript $0$ represents the functional average with the action $S_{\msf{eff}}^{(s)}|_{J=0}$ (Eq.~\ref{eq:S2stan20}).
	We find the contribution to the generating function from the standard saddle point and the quadratic fluctuations around it for zero source field $J=0$, 
	\begin{align}\label{eq:Zs-J0}
	\begin{aligned}
		\mathcal{Z}^{(s)}[J=0]
		=
		\frac{(1-\aF \bB) (1-\aB\bF)}{(1-\aF\bF)(1-\aB\bB)}.
	\end{aligned}
	\end{align}
	
	The contribution of the standard saddle point to the correlation function $\bar{C}_{nn'm'm}(\aF,\bF)$ can be obtained from derivatives of $\mathcal{Z}^{(s)}[J]$ (Eq.~\ref{eq:Seff2}) with respect to the source field:
	\begin{align}\label{eq:Cstan-0}
	\begin{aligned}
	&\bar{C}_{nn'mm'}^{(s)}(\aF,\bF)
	=
	\dfrac{\partial^2 \mathcal{Z}^{(s)}[J]}{\partial J^+_{n'n}\partial J^-_{mm'}}
	\bigg\lvert_{J=0}
	\\
	&
	=
	\mathcal{Z}^{(s)}[J=0]
	\left\langle 
	\left( 
	\delta_{nn'}\delta_{mm'}
	-
	\frac{\partial^2S_{\msf{eff}}^{(s)}}{\partial J^+_{n'n}\partial J^-_{mm'}}
	+
	\frac{\partial S_{\msf{eff}}^{(s)}}{\partial J^+_{n'n}}
	\frac{\partial S_{\msf{eff}}^{(s)}}{\partial J^-_{mm'}}
	-
	\delta_{nn'}
	\frac{\partial S_{\msf{eff}}^{(s)}}{\partial J^-_{mm'}}
	-
	\delta_{mm'}
	\frac{\partial S_{\msf{eff}}^{(s)}}{\partial J^+_{n'n}}
	\right)  \bigg\lvert_{J=0}
	\right\rangle_0. 
	\end{aligned}
	\end{align}
	Using the expressions for the quadratic action $S_{\msf{eff}}^{(s)}$ in Eq.~\ref{eq:S2stan} and the bare propagators in Eq.~\ref{eq:bareProp}, we find
	{\allowdisplaybreaks
	\begin{subequations}\label{eq:dS}
		\begin{align}
	&\begin{aligned}
	\left\langle \frac{\partial S_{\msf{eff}}^{(s)}}{\partial J^+_{n'n}}\bigg\lvert_{J=0}\right\rangle_0
	=&
	\delta_{nn'} \frac{1}{N^2} 
	\sum_{a}
	s_{a}\aF \beta_a 
	\left\langle \tilde{X}^{a\msf{F}} (0) X^{\msf{F}a} (0) \right\rangle_0 
	=
	\delta_{nn'}  \frac{1}{N}  
	\frac{\aF (\bF-\bB)}{(1-\aF \bF)(1-\aF \bB)},
	\end{aligned}
	\\
	&\begin{aligned}
	\left\langle \frac{\partial S_{\msf{eff}}^{(s)}}{\partial J^-_{mm'}}\bigg\lvert_{J=0} \right\rangle_0
	=&
	\delta_{mm'} \frac{1}{N^2}
	\sum_{b=B,F}
	s_{F}\alpha_b \bF 
	\left\langle \tilde{X}^{\msf{F}b} (0) X^{b\msf{F}} (0) \right\rangle_0
	=
	\delta_{mm'}  \frac{1}{N} 
	\frac{\bF(\aF-\aB)}{(1-\aF \bF)(1-\aB \bF)} ,
	\end{aligned}
	\\
	&
	\begin{aligned}
	\left\langle 	\frac{\partial^2 S_{\msf{eff}}^{(s)}}{\partial J^+_{n'n}\partial J^-_{mm'}}\bigg\lvert_{J=0}\right\rangle_0 
	=	&
	-
	\delta_{nn'} \delta_{mm'}
	\frac{1}{N^3} 
	\sum_{k=0}^{N-1} 	s_{F} \aF \bF 
	\left\langle \tilde{X}^{\msf{FF}} (k) X^{\msf{FF}} (k)\right\rangle_0  
	e^{-i2\pi k (n-m)/N}  
	\\
	&
	-
	(1-\delta_{nn'}) (1-\delta_{mm'}) 
	\frac{1}{N} 
	s_{F}\aF  \bF 
	\left\langle 
	\tilde{Y}_{nn'}^{\msf{FF}} 
	Y_{m'm}^{\msf{FF}} \right\rangle_0
		\\
			=&
			-
			\delta_{nn'} \delta_{mm'}
			\frac{1}{N^2} 
			\frac{(\aF \bF)^2}{1-\aF \bF }
			-
			\delta_{nm}\delta_{n'm'}
			\frac{1}{N} 
			\aF  \bF ,
	\end{aligned}
	\\
	&\begin{aligned}
	\left\langle \frac{\partial S_{\msf{eff}}^{(s)}}{\partial J^+_{n'n}}
	\frac{\partial S_{\msf{eff}}^{(s)}}{\partial J^-_{mm'}}\bigg\lvert_{J=0}
	\right\rangle_0
	=&
	\frac{1}{N^2} 
	\sum_{ab}
	s_{a}s_{F} \aF \bF \alpha_b \beta_a 
	\left\langle 
	\left( 
	\delta_{nn'} 
	\sum_{k=0}^{N-1}\tilde{X}^{a\msf{F}} (k) \frac{1}{N}  e^{-i2\pi k n/N} 
	+
	(1-\delta_{nn'}) \tilde{Y}_{nn'}^{a\msf{F}}  
	\right)	X^{\msf{F}a} (0) 
	\right. 
	\\
	&\times
	\left. 
	\tilde{X}^{\msf{F}b} (0) 
	\left( 
	\delta_{mm'} \sum_{k=0}^{N-1} X^{b\msf{F}} (k) \frac{1}{N}  e^{i2\pi k m/N}  
	+
	(1-\delta_{mm'}) Y_{m'm}^{b\msf{F}}   
	\right) 
	\right\rangle_0
	\\
	=&
	\delta_{nn'} \delta_{mm'} 
	\frac{1}{N^2}  
	\frac{\aF\bF  (\bF-\bB)(\aF-\aB)}{(1-\aF \bF)^2(1-\aF \bB)(1-\aB \bF)}
	\\
	+&
	\frac{\aF \bF}{1-\aF \bF}
	\left(
	\delta_{nn'} \delta_{mm'}
	\frac{1}{N^2} 
	\frac{(\aF \bF)^2}{1-\aF \bF }
	+
	\delta_{nm}\delta_{n'm'}
	\frac{1}{N} 
	\aF  \bF \right).
	\end{aligned}
	\end{align}
	\end{subequations}
	}

Inserting these equations into Eq.~\ref{eq:Cstan-0}, and setting $\aB=\aF$ and $\bB=\bF$, we obtain the standard saddle point's contribution to the correlation function $\bar{C}_{nn'm'm}(\aF,\bF)$:
\begin{align}
\begin{aligned}\label{eq:Cstan-1}
	\bar{C}_{nn'm'm}^{(s)}(\aF,\bF)
	=\,&
	\delta_{nn'}\delta_{mm'}
	+
	\delta_{nn'} \delta_{mm'} 
	\frac{1}{N^2} 
	\left( 
	\frac{\aF\bF}{1-\aF\bF}
	\right)^2 
	+
		\delta_{nm} \delta_{n'm'}
		\frac{1}{N} 
		\frac{\aF\bF}{1-\aF\bF}.
\end{aligned}
\end{align}
This leads to
\begin{align}\label{eq:Cstan}
\begin{aligned}
	C_{nn'm'm}^{(s)}(\omega)
	=&
	\delta_{nn'} \delta_{mm'} 
	\frac{1}{2\pi}
	\left[  
	1+
	\frac{1}{N^2} 
	\left( 
	1
	-
	\frac{1}{2}
	\frac{1}{\sin^2(\w/2)}
	\right)
	\right] 
	-
		\delta_{nm} \delta_{n'm'}
		\frac{1}{N} 
		\frac{1}{2\pi},
\end{aligned}
\end{align}
where we have used Eq.~\ref{eq:C2} and set $\aF\bF=e^{i\w}$.

As mentioned earlier, consideration of only the standard saddle point gives rise to the smoothed correlation function.
We can see that Eq.~\ref{eq:Cstan} is consistent with the RMT prediction Eq.~\ref{eq:Cw-exact}, to the leading order in $1/N$, after replacing the two-level correlation function $R_2(\w)$ with its smoothed version:
\begin{align}
\begin{aligned}
	R_2^{(s)}(\w)
	=
	\frac{N^2}{2\pi}
	\left( 
	1
	-
	\frac{1}{N^2}
	\frac{1}{2\sin^2(\w/2)}
	\right).
\end{aligned}
\end{align}
For the comparison of higher order terms in Eq.~\ref{eq:Cstan} and Eq.~\ref{eq:Cw-exact}, see the discussion at the end of this section.

	\subsection{Quadratic fluctuations around the nonstandard saddle point}
	
	To study the quadratic fluctuations around the nonstandard saddle point, we insert
	\begin{align}
		Z=Z^{(ns)}+\delta Z,
		\qquad
		\TZ=\TZ^{(ns)}+\delta \TZ,
	\end{align}
	into the action $S[\TZ,Z]$ in Eq.~\ref{eq:sigma} and expand it up to the quadratic order in $\delta Z$ and $\delta \TZ$.
	This gives rise to
	\begin{align}
	\begin{aligned}
	\delta S_2^{(ns)} [\delta \TZ,\delta Z]
	=\, &
	\STr  \left[ (1-\TZsp {\Zsp})^{-1}\delta \tilde{Z}\delta Z \right] 
	-
	\STr  \left[ (1- \TZsp \alpha_J \mathbf{U} \Zsp \beta_J \mathbf{U}^{\dagger})^{-1} \delta \tilde{Z} \alpha_J \mathbf{U}  \delta Z \beta_J \mathbf{U}^{\dagger} \right] 
	\\
	&
	+\frac{1}{2}
	\STr \left[ 
	\left(
	(1-\TZsp {\Zsp})^{-1} 
	( 
	\delta \tilde{Z}\Zsp 
	+
	\TZsp
	\delta Z 
	) 
	\right)^2
	\right] 
	\\
	&
	-\frac{1}{2}
	\STr 
	\left[ 
	\left( 
	(1- \TZsp \alpha_J \mathbf{U}  \Zsp \beta_J \mathbf{U}^{\dagger})^{-1} 
	( 
	\delta \tilde{Z} 
	\alpha_J \mathbf{U}  \Zsp \beta_J \mathbf{U}^{\dagger} 
	+
	\TZsp \alpha_J \mathbf{U} 
	\delta Z 
	\beta_J \mathbf{U}^{\dagger}
	)
	\right)^2 
	\right] .
	\end{aligned}
	\end{align}
	Making use of the expression for the nonstandard saddle point $Z^{(ns)}$ (Eq.~\ref{eq:AA}) and the saddle point equation (Eq.~\ref{eq:SPEQz}),
	we obtain 
	\begin{align}\label{eq:dsns}
	\begin{aligned}
	&\delta S_2^{(ns)} [\delta \TZ,\delta Z]
	=
	+
	\Tr  \left[ \delta \tilde{Z}^{\msf{BB}}\delta Z^{\msf{BB}} \right] 
	-
	\Tr  \left[  \delta \tilde{Z}^{\msf{BB}} \aB  U  \delta Z^{\msf{BB}} \bB U^{\dagger}  \right] 
	\\
	&-
	\Tr  \left[ 
	\delta \tilde{Z}^{\msf{FF}}
	(1+zz^{\dagger})^{-1}
	\delta Z^{\msf{FF}} 
	(1+z^{\dagger}z)^{-1}
	\right] 
	+
	\Tr 
	\left[ 
	\delta \tilde{Z}^{\msf{FF}} 
	( z^{\dagger} )^{-1} U (\beta_J^{\msf{FF}})^{-1} z^{\dagger} (1+zz^{\dagger})^{-1}
	\delta Z^{\msf{FF}}
	z^{-1} U^{\dagger} (\alpha_J^{\msf{FF}})^{-1} z
	(1+z^{\dagger}z)^{-1}
	\right] 
	\\
	&+
	\Tr  \left[ 
	\delta \tilde{Z}^{\msf{BF}}
	(1+zz^{\dagger})^{-1}
	\delta Z^{\msf{FB}} 
	\right] 
	-
	\Tr 
	\left[ 
	\delta \tilde{Z}^{\msf{BF}}
	( z^{\dagger} )^{-1} U (\beta_J^{\msf{FF}})^{-1} z^{\dagger} (1+zz^{\dagger})^{-1}
	\delta Z^{\msf{FB}}
	\bB U^{\dagger}
	\right] 
	\\
	&-
	\Tr  \left[ \delta \tilde{Z}^{\msf{FB}}\delta Z^{\msf{BF}} (1+z^{\dagger}z)^{-1} \right] 
	+
	\Tr  \left[  \delta \tilde{Z}^{\msf{FB}} \aB U \delta Z^{\msf{BF}} 
	z^{-1} U^{\dagger} (\alpha_J^{\msf{FF}})^{-1} z
	(1+z^{\dagger}z)^{-1}
	\right].
	\end{aligned}
	\end{align}
	
	Applying the transformation:
	\begin{align}
	\begin{aligned}
	 &\delta Z^{\msf{BB}} \rightarrow \delta Z^{\msf{BB}},
	 \quad
	 \delta Z^{\msf{FF}}  \rightarrow (1+zz^{\dagger}) (z^{\dagger} )^{-1}\delta Z^{\msf{FF}}z,
	 \quad
	  \delta Z^{\msf{FB}} \rightarrow (1+zz^{\dagger}) (z^{\dagger})^{-1}\delta Z^{\msf{FB}} ,
	 \quad
	 \delta Z^{\msf{BF}} \rightarrow \delta Z^{\msf{BF}}z ,
	 \\
	 & \delta \TZ^{\msf{BB}} \rightarrow \delta \TZ^{\msf{BB}},
	 \quad
	  \delta \TZ^{\msf{FF}}  \rightarrow (1+z^{\dagger} z) z^{-1}\delta \TZ^{\msf{FF}} z^{\dagger},
	 \quad
	 \delta \TZ^{\msf{BF}}  \rightarrow \delta \TZ^{\msf{BF}} z^{\dagger},
	 \quad
	 \delta \TZ^{\msf{FB}} \rightarrow (1+z^{\dagger} z) z^{-1}\delta \TZ^{\msf{FB}},
	\end{aligned}
	\end{align}
	one finds that Eq.~\ref{eq:dsns} can be further simplified to
	\begin{align}\label{eq:ds2non}
	\begin{aligned}
	&\delta S_2^{(ns)} [\delta \TZ,\delta Z]
	=
	\STr  \left[\delta \tilde{Z} \delta Z \right] 
	-
	\STr  \left[  \delta \tilde{Z} \mathbf{U} \tilde{\alpha}_J \delta Z \mathbf{U}^{\dagger}  \tilde{\beta}_J \right]. 
	\end{aligned}
	\end{align}
	Here $\tilde{\alpha}_J$ and $\tilde{\beta}_J$ are $2N\times 2N$ matrices defined as
	\begin{align}
	\begin{aligned}
	&\tilde{\alpha}_J
	=
	\begin{bmatrix}
	\aB \mathbb{1}_{\mathcal{H}}  & 0
	\\
	0 & \bF^{-1} (1 +J^-)
	\end{bmatrix}_{\msf{BF}},
	\qquad
	\tilde{\beta}_J
	=
	\begin{bmatrix}
	\bB  \mathbb{1}_{\mathcal{H}}  & 0
	\\
	0 & \aF^{-1}(1+J^+)
	\end{bmatrix}_{\msf{BF}}.
	\end{aligned}
	\end{align}
	
	Note that although $\mathbf{U}$ and  $\tilde{\alpha}_J $ ($\mathbf{U}^{\dagger} $ and $\tilde{\beta}_J$) do not always commute, one can still permute them in the action. In particular, if one apply the transformation
	\begin{align}
	\begin{aligned}
	\delta Z \rightarrow \mathbf{U} \delta Z \mathbf{U}  ,
	\qquad
	\delta \TZ  \rightarrow  \mathbf{U}^{\dagger} \delta Z \mathbf{U}^{\dagger},
	\end{aligned}
	\end{align}
	the action becomes
	\begin{align}\label{eq:ds2non2}
	\begin{aligned}
		&\delta S_2^{(ns)} [\delta Z,\delta \TZ]
		=
		\STr  \left[\delta \tilde{Z} \delta Z \right] 
		-
		\STr  \left[  \delta \tilde{Z} \tilde{\alpha}_J \mathbf{U} \delta Z \tilde{\beta}_J \mathbf{U}^{\dagger}   \right] .
	\end{aligned}
	\end{align}
	which is identical to that of the standard saddle point (Eq.~\ref{eq:S2stan}) after the replacement $\alpha_J \rightarrow \tilde{\alpha}_J$	 and $\beta_J \rightarrow \tilde{\beta}_J$.
	
	The contribution of the nonstandard saddle point to the generating function can be approximated as
	\begin{align}\label{eq:Zns}
	\begin{aligned}
		\mathcal{Z}^{(ns)}[J]
		=
		(\aF\bF)^{N}
		\int \D(\TZ,Z) e^{- S_{\msf{eff}}^{(ns)}},
		\qquad
		S_{\msf{eff}}^{(ns)}=\left\langle \delta S_2^{(ns)} [ \TZ, Z]\right\rangle,
	\end{aligned}
	\end{align}
	where we have used the expression for the action at the nonstandard saddle point Eq.~\ref{eq:Ssaddle-ns}.
	Given the close similarity between their actions (Eq.~\ref{eq:S2stan-0} and Eq.~\ref{eq:ds2non2}), the contribution from the quadratic fluctuations around the nonstandard saddle point can be deduced from that of the standard saddle point by making the replacement.
	\begin{align}\label{eq:replacement}
	\begin{aligned}
		{\aF\rightarrow \bF^{-1}, 
		\quad
		\bF\rightarrow \aF^{-1}, 
		\quad
		J^+ \rightarrow (1+J^-)^{-1}-1\approx -J^-,
		\quad
		 J^- \rightarrow (1+J^+)^{-1}-1}\approx -J^+.
	\end{aligned}
	\end{align}

From Eq.~\ref{eq:Zns}, we find that the contribution from the nonstandard saddle point and its quadratic fluctuations to the correlation function $\bar{C}_{nn'm'm}(\w)$ can be expressed as 
	\begin{align}\label{eq:Cnonstan-0}
\begin{aligned}
&\bar{C}_{nn'mm'}^{(ns)}(\omega)
=
\dfrac{\partial^2 \mathcal{Z}^{(ns)}[J]}{\partial J^+_{n'n}\partial J^-_{mm'}}
\bigg\lvert_{J=0}
=
\mathcal{Z}^{(ns)}[J=0]
\left\langle 
\left( 
-
\frac{\partial^2S_{\msf{eff}}^{(ns)}}{\partial J^+_{n'n}\partial J^-_{mm'}}
+
\frac{\partial S_{\msf{eff}}^{(ns)}}{\partial J^+_{n'n}}
\frac{\partial S_{\msf{eff}}^{(ns)}}{\partial J^-_{mm'}}
\right)\bigg\lvert_{J=0}
\right\rangle_0. 
\end{aligned}
\end{align}
	Here $Z^{(ns)}[J=0]/(\aF\bF)^{N}$ and the derivatives of the ensemble averaged effective action $S_{\msf{eff}}^{(ns)}$ with respect to the source field $J^{\pm}$ can be deduced from their standard saddle point counterparts (Eqs.~\ref{eq:Zs-J0} and~\ref{eq:dS}) by applying the replacement Eq.~\ref{eq:replacement}:
		\begin{subequations}\label{eq:dS-ns}
		\begin{align}
		& \begin{aligned}
		\mathcal{Z}^{(ns)}[J=0]
		=
		-
		(\aF\bF)^N
		\dfrac{(\aF-\aB)(\bF-\bB)}{(1-\aB\bB)(1-\aF\bF)},
		\end{aligned}
		\\
		&
		\begin{aligned}
		\left\langle 	\frac{\partial^2 S_{\msf{eff}}^{(ns)}}{\partial J^+_{n'n}\partial J^-_{mm'}}\bigg\lvert_{J=0}\right\rangle_0 
		=&
		\delta_{nn'} \delta_{mm'}
		\frac{1}{N^2} 
		\frac{1}{(1-\aF \bF)\aF\bF}
		-
		\delta_{nm}\delta_{n'm'}
		\frac{1}{N} 
		\frac{1}{\aF  \bF},
		\end{aligned}
		\\
		&\begin{aligned}
		\left\langle \frac{\partial S_{\msf{eff}}^{(ns)}}{\partial J^+_{n'n}}
		\frac{\partial S_{\msf{eff}}^{(ns)}}{\partial J^-_{mm'}}\bigg\lvert_{J=0}
		\right\rangle_0
		=&
		\delta_{nn'} \delta_{mm'} 
		\frac{1}{N^2}  
		\frac{(1-\aF \bB)(1-\aB \bF)\aF\bF}{(1-\aF \bF)^2(\aF -\aB)(\bF-\bB)}
		\\
		-&
		\frac{1}{1-\aF \bF}
		\left(
		-\delta_{nn'} \delta_{mm'}
		\frac{1}{N^2} 
		\frac{1}{(1-\aF \bF)\aF\bF}
		+
		\delta_{nm}\delta_{n'm'}
		\frac{1}{N} 
		\frac{1}{\aF\bF}
		\right).
		\end{aligned}
		\end{align}
		\end{subequations}
	 	Inserting these equations into Eq.~\ref{eq:Cnonstan-0} and setting $\aB=\aF$ and $\bB=\bF$, we obtain
	 	\begin{align}\label{eq:Cnonstan-1}
	 	\begin{aligned}
	 	&\bar{C}_{nn'mm'}^{(ns)}(\aF,\bF)
	 	=
	 	-\delta_{nn'} \delta_{mm'} 
		\frac{1}{N^2}  
		(\aF\bF)^N
		\frac{\aF\bF}{(1-\aF \bF)^2}.
		\end{aligned}
		\end{align}
		

Adding this result with the contribution of the standard saddle point (Eq.~\ref{eq:Cstan-1}), we find the correlation function $\bar{C}_{nn'm'm}(\aF,\bF)$:
	\begin{align}
	\begin{aligned}\label{eq:C-1}
	\bar{C}_{nn'm'm}(\aF,\bF)
	=\,&
	\delta_{nn'} \delta_{mm'} 
	\left[ 
	1+
	\frac{1}{N^2} 
	\left( 1- (\aF\bF)^N\right) 
	\frac{\aF\bF}{(1-\aF\bF)^2}
	-
	\frac{1}{N^2} 
	\frac{\aF\bF}{1-\aF\bF}
	\right] 
	+
	\delta_{nm} \delta_{n'm'}
	\frac{1}{N} 
	\frac{\aF\bF}{1-\aF\bF}.
	\end{aligned}
	\end{align}
	which leads to
 	\begin{align}\label{eq:Cw-final}
	\begin{aligned}
	C_{nn'm'm}(\w)
	=&
	\left[ 
	\frac{1}{2\pi}
	-
	\frac{1}{N^2} 
	\frac{1}{2\pi}
	\dfrac{\sin^2 (N \w/2)}{\sin^2 (\w/2)}
	+
	\frac{1}{N}
	\delta(\w)
    +
    \frac{1}{N^2} 
	\frac{1}{2\pi}
	\right] 
	\delta_{nn'}\delta_{mm'}
		-
	\frac{1}{N} 
	\frac{1}{2\pi}
	\delta_{nm}\delta_{n'm'}
	+
	\\
	&
		-
		\frac{1}{N^2} 
		\delta(\w)
			\delta_{nn'} \delta_{mm'} 
		+
		\frac{1}{N}
		\delta(\w)
		\delta_{nm}\delta_{n'm'}
		,
\end{aligned}
\end{align}
Here we have made use of Eq.~\ref{eq:C2} and the following identity:
\begin{align}
\begin{aligned}
2\re \frac{e^{i\w}}{1-e^{i\w}} +1
=2\pi\delta(\w).
\end{aligned}
\end{align}

Eq.~\ref{eq:Cw-final} is consistent with the RMT prediction Eq.~\ref{eq:Cw-exact} to the leading order in $1/N$.
In particular, we have obtained the additional oscillatory term 
$\frac{1}{4 \pi N^2}\frac{\cos (N \w)}{\sin^2 (\w/2)}\delta_{nn'}\delta_{mm'}$
by taking into account the contribution from the nonstandard saddle point.
Note that the higher order terms in Eq.~\ref{eq:Cw-final} do not agree with the corresponding terms in Eq.~\ref{eq:Cw-exact}.
In particular, instead of 
$\delta_{nn'} \delta_{mm'} R_2(\w)/N^4$
and
$-\delta_{nm}\delta_{n'm'}R_2(\w)/N^3$
in the CUE result Eq.~\ref{eq:Cw-exact},
we find
$\delta_{nn'} \delta_{mm'} R_2^{(dis)}(\w)/N^4$
and
$-\delta_{nm}\delta_{n'm'}R_2^{(dis)}(\w)/N^3$,
where the disconnected part of the two-level correlation function is $R_2^{(dis)}(\w)=N^2/2\pi$.
However, the higher order fluctuations are needed to find the exact result for $C_{nn'm'm}(\w)$ at these higher orders and to compare with the RMT prediction.
The calculation of the contribution from the higher order fluctuations is rather cumbersome.
We instead consider for simplicity the CUE in Sec.~\ref{sec:HOF}, and investigate the higher order fluctuations of the supermatrix field $Z$ in its sigma model.
We demonstrate that the next leading higher order term $-\delta_{mn} \delta_{n'm'}R_2(\w)/N^3$ in the CUE result can indeed be recovered within the sigma model approach by taking into account the higher order fluctuations.

	
	\section{Higher order fluctuations for the CUE}\label{sec:HOF}
	
	In this section, we use the SUSY sigma model to rederive the correlation function $C_{nn'm'm}(\w)$ for the CUE, i.e., Eq.~\ref{eq:Cw-exact}.
		The goal of this section is to demonstrate that the second term in Eq.~\ref{eq:Cw-exact}, which is of higher order in the large $N$ expansion, arises from the higher order fluctuations. 
	Unlike the earlier derivation in Sec.~\ref{sec:RMT} which only shows the relation between $C_{nn'm'm}(\w)$ and $R_2(\w)$, this sigma model calculation also gives the explicit expression for $R_2(\w)$ (Eq.~\ref{eq:R2}).
	Moreover, there is no need of the Weingarten calculus which can also be derived in the sigma model framework~\cite{entropy}.
    Note that the sigma model calculation performed in Sec.~\ref{sec:RQC} also applies directly to the CUE whose second moment of the Floquet operator is also given by Eq.~\ref{eq:U2}.
	In this section, we employ a slightly different approach which makes use of the color-flavor transformation to perform the ensemble averaging.
	This approach greatly simplifies the structure of the sigma model (see details below), but can not be directly applied to the Floquet random quantum circuits we considered.

	\subsection{Sigma model for the CUE}
	
	We start from the superintegral representation of the generating function $\mathcal{Z}[J]$ Eq.~\ref{eq:ZJ-1}, which is applicable to an arbitrary ensemble of Floquet systems.
	Instead of converting the integration over the center phase $\phi$ into an integration over supermatrix matrices,
	we first average over the ensemble of the CUE matrices $U$. 
	This ensemble averaging can be performed
	using the color-flavor transformation~\cite{Zirnbauer_1996,Zirnbauer_1998,Zirnbauer_1999,Zirnbauer_2021}:
	 \begin{align}\label{eq:ZJ-cue-0}
	\begin{aligned}
	\mathcal{Z}[J]
	=&
	\det(1+J^+) \det(1+J^-)
	\int_{0}^{2\pi} \frac{d \phi}{2\pi}
	\int D(\tilde{Z},Z)
	\exp \left[ N \Str \ln (1-\tilde{Z} {Z})\right] 
	\\
	& \times
	\int
	\D (\bpsi,\psi)
	\exp
	\left\lbrace 
	\begin{aligned}
	-\begin{bmatrix}
	\bar{\psi}^{+}
	&
	\bar{\psi}^{-}
	\end{bmatrix}
	\begin{bmatrix}
	1 & e^{i\phi}\alpha_J Z 
	\\
	e^{-i\phi}\beta_J \tilde{Z} & 1
	\end{bmatrix}_{\msf{RA}}
	\begin{bmatrix}
	{\psi}^{+}
	\\
	{\psi}^{-}
	\end{bmatrix}
	\end{aligned}
	\right\rbrace.
	\end{aligned}
	\end{align}
	Here $\Str$ traces over the $\msf{BF}$ space only, in comparison with $\STr$ which is introduced earlier and traces over the product space $\msf{BF}\otimes \mathcal{H}$.
	We emphasize that, unlike in Sec.~\ref{sec:RQC}, here $Z$ and $\tilde{Z}$ are $2\times 2$ matrices defined in the $\msf{BF}$ space only.
	They are subject to the constraints $\tilde{Z}^{\msf{FF}}=-Z^{\msf{FF}}\,^{*}$, $\tilde{Z}^{\msf{BB}}=Z^{\msf{BB}}\,^{*}$, and $|Z^{\msf{BB}}|<1$.

	Integrating out the supervectors $\bpsi$ and $\psi$, we arrive at the zero-dimensional SUSY sigma model for the CUE~\cite{Zirnbauer_1996,Haake}:
	\begin{align}\label{eq:sigma0}
	\begin{aligned}
	\mathcal{Z}[J]
	=&
	\det(1+J^+) \det(1+J^-)
	\int D(\tilde{Z}, Z)
	\exp(-S[\TZ,Z]),
	\\
	S[\TZ,Z]
	=&
	-
	N\Str \ln (1-\tilde{Z} {Z})
	+
	\STr \ln (1- (\tilde{Z} \otimes \mathbb{1}_{\mathcal{H}})
	\alpha_J   (Z\otimes \mathbb{1}_{\mathcal{H}}) \beta_J).
	\end{aligned}
	\end{align}
	Compared with the sigma model in Eq.~\ref{eq:sigma} for an arbitrary ensemble of Floquet systems, here we have converted the averaging over the CUE matrices $U$ (instead of the averaging over the center phase $\phi$) into an integration over the supermatrices $Z$ and $\tilde{Z}$ which have no structure in the Hilbert space.
	It is therefore much easier to study the role played by the higher order fluctuations in the case of the CUE.

	\subsection{Saddle points}
		
	From Eq.~\ref{eq:sigma0}, we obtain the saddle point equation:
	
	\begin{align}\label{eq:SPEQ0}
	\begin{aligned}
	\left( Z(1-\tilde{Z} {Z})^{-1} \right) \otimes \mathbb{1}_{\mathcal{H}}
	=
	\alpha_J (Z\otimes \mathbb{1}_{\mathcal{H}}) \beta_J  
	\left[ 1-  (\tilde{Z} \otimes \mathbb{1}_{\mathcal{H}}) \alpha_J  (Z\otimes \mathbb{1}_{\mathcal{H}}) \beta_J \right]^{-1}.
	\end{aligned}
	\end{align}
	A trivial solution to this equation is the standard saddle point given by Eq.~\ref{eq:stan}.
	We emphasize again that $Z^{(s)}$ and $\TZ^{(s)}$ are now $2\times 2$ supermatrices acting in the $\msf{BF}$ space only.
	As before, the saddle point equation has another solution - the nonstandard saddle point. It acquires the form of Eq.~\ref{eq:AA} where $z$ is no longer a matrix in the Hilbert space but rather a complex number with$|z| =\infty$.
	The actions at the standard and nonstandard saddle points are given by Eq.~\ref{eq:Ssaddle-s} and Eq.~\ref{eq:Ssaddle-ns}, respectively.
	
	For simplicity, we will focus on the fluctuations around the standard saddle point, and derive the smoothed correlation function $C_{nn'm'm}^{(s)}(\w)$. We will show that the higher order term $-\delta_{mn} \delta_{n'm'}R_2^{(s)}(\w)/N(N^2-1)$ in the smoothed correlation function $C_{nn'm'm}^{(s)}(\w)$ for the CUE, missing from the earlier sigma model calculation, comes from higher order fluctuations around the standard saddle point.
	
	\subsection{Quadratic fluctuations around the standard saddle point}
	
	We consider first the quadratic fluctuations around  the standard saddle point $Z^{(s)}=0$ which are governed by the action:
	\begin{align}\label{eq:ds2-cue}
	\begin{aligned}
	\delta S_2^{(s)}[\TZ,Z]
	=
	N\sum_{ab}
	s_a
	\tilde{Z}^{ab} {Z}^{ba}
	\left[ 
	1
	-
	\frac{1}{N}
	\Tr(\alpha_J^{bb}\beta_J^{aa})
	\right].
	\end{aligned}
	\end{align}
	From this equation, we find the bare propagator for $Z$ arising from the quadratic action at $J=0$:
	 \begin{align}\label{eq:Zprop} 
	\begin{aligned}
	\left\langle Z^{ba} \TZ^{a'b'} \right\rangle_{0}
	=
	s_a s_b \left\langle \TZ^{a'b'} Z^{ba}  \right\rangle_{0}
	=
	\delta_{aa'}\delta_{bb'}
	\dfrac{1}{N}
	\dfrac{s_a}{1-\alpha_b \beta_a}.
	\end{aligned}
	\end{align}
	As before, we use the angular bracket with subscript $0$ to denote the functional averaging with the quadratic action $\delta S_2^{(s)}$ at  $J=0$.
	
	Performing the Gaussian integration governed by the action $\delta S_2^{(s)}$ in Eq.~\ref{eq:ds2-cue},  and making use of Eq.~\ref{eq:Ssaddle-s}, we find the contribution to the generating function from the quadratic fluctuations around the standard saddle point
	\begin{align}
	\begin{aligned}
	\mathcal{Z}^{(s)}[J]
	=&
	\det(1+J^+)\det(1+J^-)
	\dfrac{
		\left[  1-\frac{1}{N}\aB\bF \Tr\left( (1+J^-)^{-1}\right) \right] 
		\left[  1-\frac{1}{N}\aF\bB \Tr \left( (1+J^+)^{-1}\right) \right] 
	}
	{
		\left( 1-\aB\bB \right)
		 \left[ 1-\frac{1}{N}\aF\bF 
		 \Tr\left( (1+J^+)^{-1} (1+J^-)^{-1} \right)
		 	\right] }.
	\end{aligned}
	\end{align}
	Taking derivatives of the generating function $\mathcal{Z}^{(s)}[J]$ in the equation above  with respect to the source field $J^{\pm}$ and setting $\aB=\aF$ and $\bB=\bF$, we find that the quadratic fluctuations' contribution to $\bar{C}_{nn'm'm}^{(s)} (\aF,\bF)$ is given by an expression identical to the earlier result Eq.~\ref{eq:Cstan-1}, as expected.
		This also gives rise to the leading order smoothed correlation function $C^{(s)}_{nn'm'm}(\w)$ in Eq.~\ref{eq:Cstan}.
	
	\subsection{Higher order fluctuations around the standard saddle point}
	
%
	We now consider the higher order fluctuations and expand the action $S[\TZ,Z]$  around the standard saddle point up to sixth order in $Z$ and $\TZ$:
	\begin{align}\label{eq:S-cue}
	\begin{aligned}
	\delta S_6[\TZ,Z]
	=&
	\sum_{k=1}^{3}
	\frac{1}{k}
	\left\lbrace 
	N
	\Str \left[ \left( \tilde{Z} {Z}\right)^{k}  \right] 
	-
	\STr \left[ 
	\left(  (\tilde{Z} \otimes \mathbb{1}_{\mathcal{H}}) \alpha_J  (Z\otimes \mathbb{1}_{\mathcal{H}}) \beta_J \right)^{k} \right]
	\right\rbrace .
	\end{aligned}
	\end{align} 
	From this equation, it is easier to see that the correlation function $\bar{C}_{nn'm'm}$ given by the derivatives of the generating function $\mathcal{Z}[J]$ with respect $J^+_{n'n}$ and $J^-_{mm'}$ (Eq.~\ref{eq:C1}) contains two parts $\bar{C}_{nn'm'm}^{(1)}$ and $\bar{C}_{nn'm'm}^{(2)}$ which are proportional to $\delta_{nn'}\delta_{mm'}$ and  $\delta_{nm}\delta_{n'm'}$, respectively.
	In this section, we will focus on the part $\bar{C}_{nn'm'm}^{(2)}$ and recover the higher order term $-\delta_{mn} \delta_{n'm'}R_2(\w)/N^3$ in the CUE result Eq.~\ref{eq:Cw-exact}.

	From Eq.~\ref{eq:S-cue}, one can see that $\bar{C}_{nn'm'm}^{(2)}$ arises from
	\begin{align}\label{eq:C2-cue}
	\begin{aligned}
		\bar{C}_{nn'm'm}^{(2)}
		=
		 \sum_{k=1}^{3}
		  \frac{1}{k} 
		\left\langle 
		\frac{\partial^2 	\STr \left[ 
				\left(  (\tilde{Z} \otimes \mathbb{1}_{\mathcal{H}}) \alpha_J  (Z\otimes \mathbb{1}_{\mathcal{H}}) \beta_J \right)^{k} \right]
			}{\partial J_{n'n}^+\partial J_{mm'}^-}\bigg|_{J=0}
		\right\rangle.
	\end{aligned}
	\end{align}
	Here and throughout this section we use the angular bracket without any subscript (with subscript $0$) to denote the functional averaging with the action $\delta S_6[\TZ,Z]$ in Eq.~\ref{eq:S-cue} (quadratic action $\delta S_2[\TZ,Z]$ in Eq.~\ref{eq:ds2-cue}) at $J=0$.

	The $k=1$ term in Eq.~\ref{eq:C2-cue} is given by
	\begin{align}\label{eq:k=1}
	\begin{aligned}
	&
	\left\langle 
	\frac{\partial^2 	\STr \left[ 
		\left(  (\tilde{Z} \otimes \mathbb{1}_{\mathcal{H}}) \alpha_J  (Z\otimes \mathbb{1}_{\mathcal{H}}) \beta_J \right) \right]
	}{\partial J_{n'n}^+\partial J_{mm'}^-}\bigg|_{J=0}
	\right\rangle
	=
	\delta_{nm}\delta_{n'm'}
	s_F \aF \bF
	\left\langle 
	\tilde{Z}^{\msf{FF}}   Z^{\msf{FF}} 
	\right\rangle 
	\\
	&=
	\delta_{nm}\delta_{n'm'}
	s_F \aF \bF
	\left\langle 
	\tilde{Z}^{\msf{FF}}   Z^{\msf{FF}} 
	\left(
	1
	-
	S_{\msf{int}}^{(4)}
	-
	S_{\msf{int}}^{(6)}
	+
		\frac{1}{2}
		\left( 
		S_{\msf{int}}^{(4)}
		\right)^2 
	\right)
	\right\rangle_0 
	+O(1/N^4)
	\\
	&=
	\delta_{nm}\delta_{n'm'}
	\left\lbrace 
	\frac{1}{N}
	\frac{\aF \bF}{1-\aF \bF}
	-
	\frac{1}{N^3}
	\frac{\aF \bF\left[ 1-(\aF\bF)^3\right] }{(1-\aF \bF)^4}
	+
		\frac{1}{N^3}
		\frac{\aF \bF\left[ 1-(\aF\bF)^2\right]^2 }{(1-\aF \bF)^5
	}
	\right\rbrace 
	+O(1/N^4).
	\end{aligned}
	\end{align} 
	Here $S_{\msf{int}}^{(4)}$ and $S_{\msf{int}}^{(6)}$ are, respectively, the quartic order ($k=2$) and sixth order ($k=3$) term in $\delta S_6[\TZ,Z]$ (Eq.~\ref{eq:S-cue}) at $J=0$:
	\begin{align}
	\begin{aligned}
		S_{\msf{int}}^{(4)}
		=\,&
		\frac{N}{2}
		\sum_{a_1,b_1,a_2,b_2}
		s_{a_1} \TZ^{a_1 b_1} Z^{b_1 a_2}  \TZ^{a_2b_2} Z^{b_2a_1} 
		\left( 1-\alpha_{b_1} \beta_{a_2}\alpha_{b_2} \beta_{a_1}\right),
		\\
		S_{\msf{int}}^{(6)}
		=\,&
		\frac{N}{3}
		\sum_{a_1,b_1,a_2,b_2,a_3,b_3}
		s_{a_1}\TZ^{a_1 b_1} Z^{b_1 a_2}  \TZ^{a_2b_2} Z^{b_2a_3}  \TZ^{a_3b_3} Z^{b_3a_1} 
		\left( 1-\alpha_{b_1} \beta_{a_2}\alpha_{b_2} \beta_{a_3}  \alpha_{b_3}\beta_{a_1} \right).
	\end{aligned} 
	\end{align}
	 In the last equality of Eq.~\ref{eq:k=1}, we have used the expression for the bare $Z$ propagator at $J=0$ (Eq.~\ref{eq:Zprop}) and set $\aB=\aF$ and $\bB=\bF$.
	 
	Similarly, we find the $k=2$ term in Eq.~\ref{eq:C2-cue} :
	\begin{align}
	\begin{aligned}
	&\frac{1}{2}
		\left\langle 
	\frac{\partial^2 	\STr \left[ 
		\left(  (\tilde{Z} \otimes \mathbb{1}_{\mathcal{H}}) \alpha_J  (Z\otimes \mathbb{1}_{\mathcal{H}}) \beta_J \right)^{2} \right]
	}{\partial J_{n'n}^+\partial J_{mm'}^-}\bigg|_{J=0}
	\right\rangle
	\\
	&=
	\delta_{nm}\delta_{n'm'}
	\sum_{ab} 
	\left\langle
	s_a \tilde{Z}^{aF} \aF  Z^{\msf{FF}} \bF
	\tilde{Z}^{Fb} \alpha_{b} Z^{ba} \beta_a
	+
	s_F \tilde{Z}^{\msf{FF}} \aF  Z^{Fa} \beta_a
	\tilde{Z}^{ab} \alpha_{b} Z^{bF} \bF
	\right\rangle 
	\\
	&=
	\delta_{nm}\delta_{n'm'}
	\left\langle 
	\sum_{ab} 
	\left(
		s_a \tilde{Z}^{aF} \aF  Z^{\msf{FF}} \bF
	\tilde{Z}^{Fb} \alpha_{b} Z^{ba} \beta_a
	+
	s_F \tilde{Z}^{\msf{FF}} \aF  Z^{Fa} \beta_a
	\tilde{Z}^{ab} \alpha_{b} Z^{bF} \bF
	\right)
	\left(1
	-
	S_{\msf{int}}^{(4)}
	\right)
	\right\rangle_0 
	+O(1/N^4)
	\\
	&=
	-
	\delta_{nm}\delta_{n'm'}
	\frac{2}{N^3}
	\frac{(\aF \bF)^2\left[ 1-(\aF\bF)^2\right] }{(1-\aF \bF)^4}
	+O(1/N^4),
	\end{aligned}
	\end{align} 
	as well as the $k=3$ term:
	\begin{align}
	\begin{aligned}
	&\frac{1}{3} 
	\left\langle 
	\frac{\partial^2 	\STr \left[ 
		\left(  (\tilde{Z} \otimes \mathbb{1}_{\mathcal{H}}) \alpha_J  (Z\otimes \mathbb{1}_{\mathcal{H}}) \beta_J \right)^{3} \right]
	}{\partial J_{n'n}^+\partial J_{mm'}^-}\bigg|_{J=0}
	\right\rangle
	\\
	&=
	\delta_{nm}\delta_{n'm}
	\sum_{aba'b'}
	\left[ 
	s_a \left\langle \TZ^{aF} \aF Z^{\msf{FF}} \bF \TZ^{Fb} \alpha_b Z^{ba'} \beta_{a'} \TZ^{a'b'} \alpha_{b'} Z^{b'a} \beta_a\right\rangle_0
	\right. 
	\\
	&
	\left. 
	+
	s_a
	\left\langle \TZ^{aF} \aF Z^{Fb} \beta_b \TZ^{ba'} \alpha_{a'} Z^{a'F} \bF \TZ^{Fb'} \alpha_{b'} Z^{b'a} \beta_a\right\rangle_0
	+
	s_F
	\left\langle \TZ^{\msf{FF}} \aF Z^{Fa} \beta_a \TZ^{ab} \alpha_b Z^{ba'} \beta_{a'} \TZ^{a'b'} \alpha_{b'} Z^{b'F} \bF\right\rangle_0
	\right] 
	+O(1/N^4)
	\\
	&=
	\delta_{nm}\delta_{n'm}
	\frac{3}{N^3}
	\left( \frac{\aF\bF}{1-\aF\bF} \right)^3 
	+O(1/N^4).
	\end{aligned}
	\end{align}

	Combining everything, we arrive at the contribution from the first few leading order fluctuations around the standard saddle point to $\bar{C}_{nn'm'm}^{(2)}$ (i.e., the $\delta_{nm}\delta_{n'm'}$ proportional term in $\bar{C}^{(s)}_{nn'm'm}$):
			\begin{align}
			\begin{aligned}
			& \bar{C}_{nn'm'm}^{(2)} (\aF,\bF)
			=
			\delta_{nm}\delta_{n'm'}
			\left[ 
			\frac{1}{N}
			\frac{\aF \bF}{1-\aF \bF}
			-
			\frac{1}{N^3}
			\left( \frac{\aF\bF}{1-\aF\bF}\right)^2 
			\right].
			\end{aligned}
			\end{align}
		Note that the leading order first term in the equation above originates from the quadratic fluctuations (equivalent to the last term in Eq.~\ref{eq:Cstan-1}), while the higher order second term arises from the higher order fluctuations.
		It is straightforward to see that the corresponding contribution to smoothed correlation function $C_{nn'm'm}^{(s)}$ is given by
		\begin{align}
		\begin{aligned}
		 C^{(s)}_{nn'm'm}(\w)
		=&
		-\delta_{nm}\delta_{n'm'}
		\left[ 
		\frac{1}{N}\frac{1}{2\pi}
		+
		\frac{1}{N^3 }\frac{1}{2\pi} \left( 1-\frac{1}{2\sin^2 (\w/2)}\right) 
		\right],
		\end{aligned}
		\end{align}
		We have therefore reproduced the higher order term $\delta_{mn} \delta_{n'm'} R_2^{(s)}(\w)/N^3 $ in the smoothed correlation function $C_{nn'm'm}^{(s)}(\w)$ for the CUE  by inclusion of higher order fluctuations.

	
\section{Connection to physical properties}~\label{sec:PP}

Once the correlation function $C_{nn'm'm} (\w)$ defined in Eq.~\ref{eq:Cw} is known, one can immediately obtain many important physical properties of the system.
In this section, we use our result for the correlation function $C_{nn'm'm} (\w)$ in Eq.~\ref{eq:Cw-final} to evaluate the SFF,  the PSFF and the dynamical correlation function of operators, for the Floquet random quantum circuits under consideration.

\subsection{Spectral form factor}

From their definitions Eqs.~\ref{eq:Ct} and~\ref{eq:Kt}, one can see that the SFF $K(t)$ is related to $C_{nn'm'm}(t)$ by
\begin{align}\label{eq:Kt-0}
\begin{aligned}
	K(t)=\sum_{nm} C_{nnmm}(t).
\end{aligned}
\end{align}
This is consistent with the RMT prediction Eq.~\ref{eq:Ct-exact}. In particular, if we substitute Eq.~\ref{eq:Ct-exact} into the right-hand-side (RHS) of the equation above, one can immediate see that it reduces to 
\begin{align}\label{eq:Kt-11}
\begin{aligned}
\sum_{nm} C_{nnmm}(t)
=
\sum_{nm} 
\left( 
\frac{K(t)-1}{N^2-1} 
+
\frac{-K(t)/N+N}{N^2-1} 
\delta_{mn} 
\right) 
=
K(t)
,
\end{aligned}
\end{align}
identical to the left-hand-side (LHS) of Eq.~\ref{eq:Kt-0}.

Using our leading order in $1/N$ result for $C_{nn'm'm}$ in Eq.~\ref{eq:Cw-final}, we find
\begin{align}\label{eq:Kt-12}
\begin{aligned}
\sum_{nm} C_{nnmm}(t)
\approx 
\sum_{nm} 
\left( 
\dfrac{K(t)-1+\delta_{t,0}}{N^2} 
+
\dfrac{-\delta_{t,0}+1}{N} 
\delta_{mn} 
\right) 
=
K(t)
.
\end{aligned}
\end{align}
 Here $K(t)$ on the RHS 
 is the CUE SFF in Eq.~\ref{eq:Kt-cue}.
Eq.~\ref{eq:Kt-12} shows that the SFF for the Floquet random quantum circuits under study agrees with the RMT result Eq.~\ref{eq:Kt-cue}.
Note that the difference between the first equalities of Eq.~\ref{eq:Kt-11} and Eq.~\ref{eq:Kt-12} arises from the higher order terms in $C_{nnmm}(t)$, which are not important for the leading order SFF.

\subsection{Partial spectral form factor}

The partial spectral form factor (PSFF) is defined as~\cite{Zoller,Chalker-4}
\begin{align}
	K_{\mathcal{A}}(t)=\left\langle \Tr_{\bar{\mathcal{A}}} \left[ \Tr_\mathcal{A} U(t) \Tr_\mathcal{A} U^{\dagger}(t) \right] \right\rangle, 
\end{align}
where
$\Tr_{\mathcal{A}}$ and $\Tr_{\bar{\mathcal{A}}}$ represent the trace operators over the Hilbert spaces of the subsystem $\mathcal{A}$ and its complement $\bar{\mathcal{A}}$, respectively.
In terms of $C_{nn'm'm}(t)$, PSFF can be expressed as
\begin{align}\label{eq:PSFF-0}
\begin{aligned}
K_A (t)
=
\sum_{i_1^{\mathcal{A}},i_2^{\mathcal{A}}=1}^{N_{\mathcal{A}}}
\sum_{j_1^{\bar{\mathcal{A}}},j_2^{\bar{\mathcal{A}}}=1}^{N/N_{\mathcal{A}}}
C_{(i_1^{A}, j_1^{\bar{\mathcal{A}}}), (i_1^{A}, j_2^{\bar{\mathcal{A}}}),(i_2^{A}, j_2^{\bar{\mathcal{A}}}),(i_2^{A}, j_1^{\bar{\mathcal{A}}})}(t)
,
\end{aligned}
\end{align}
with $N_{\mathcal{A}}$ being the dimension of the Hilbert space of subsystem $\mathcal{A}$.
Here we have used a two-component vector $(i^{A}, j^{\bar{\mathcal{A}}})$ to label the Hilbert space of the entire system, whose first and second components label the subsystem $\mathcal{A}$ and its complement $\bar{\mathcal{A}}$, respectively.
 
Inserting the expression for the CUE correlation function $C_{nn'm'm}(t)$ Eq.~\ref{eq:Ct-exact} into the equation above, we find the RMT prediction for PSFF:
\begin{align}\label{eq:PSFF-1}
\begin{aligned}
K_{\mathcal{A}}(t) 
=&
\sum_{i_1^{\mathcal{A}},i_2^{\mathcal{A}}=1}^{N_{\mathcal{A}}}
\sum_{j_1^{\bar{\mathcal{A}}},j_2^{\bar{\mathcal{A}}}=1}^{N/N_{\mathcal{A}}}
\left(
\frac{K(t)-1}{N^2-1} 
\delta_{j_1^{\bar{\mathcal{A}}},j_2^{\bar{\mathcal{A}}}}
+
\frac{-K(t)/N+N}{N^2-1} 
 \delta_{i_1^{\mathcal{A}},i_2^{\mathcal{A}}}
\right)
=
\frac{K(t)-1}{N^2-1} 
N N_{\mathcal{A}}
+
\frac{-K(t)/N+N}{N^2-1} 
\frac{N^2}{N_{\mathcal{A}}}
\\
=&
\frac{NN_{\mathcal{A}}-N/N_{\mathcal{A}}}{N^2-1} 
K(t)
+
\frac{N^3/N_{\mathcal{A}}-NN_{\mathcal{A}}}{N^2-1},
\end{aligned}
\end{align}
This equation is equivalent to the result presented in Ref.~\cite{Zoller} as expected.

On the other hand, our leading order result for $C_{nn'm'm}$ Eq.~\ref{eq:Cw-final} leads to
\begin{align}\label{eq:PSFF-2}
\begin{aligned}
K_{\mathcal{A}}(t) 
\approx&
\sum_{i_1^{\mathcal{A}},i_2^{\mathcal{A}}=1}^{N_{\mathcal{A}}}
\sum_{j_1^{\bar{\mathcal{A}}},j_2^{\bar{\mathcal{A}}}=1}^{N/N_{\mathcal{A}}}
\left( 
\frac{K(t)-1+\delta_{t,0}}{N^2} 
\delta_{j_1^{\bar{\mathcal{A}}},j_2^{\bar{\mathcal{A}}}}
+
\frac{-\delta_{t,0}+1}{N} 
\delta_{i_1^{\mathcal{A}},i_2^{\mathcal{A}}}
\right) 
=
\frac{K(t)-1+\delta_{t,0}}{N^2} 
NN_{\mathcal{A}}
+
\frac{-\delta_{t,0}+1}{N} 
\frac{N^2}{N_{\mathcal{A}}}
\\
= &
\frac{N_{\mathcal{A}}}{N} 
K(t)
+
\left( \frac{N}{N_{\mathcal{A}}}
-
\frac{N_{\mathcal{A}}}{N}\right) 
(1-\delta_{t,0}).
\end{aligned}
\end{align}
This is consistent with the RMT result Eq.~\ref{eq:PSFF-1} when $N_{\mathcal{A}}$ the Hilbert space dimension of the subsystem $\mathcal{A}$ satisfies $ N_{\mathcal{A}}\gg 1$.

We emphasize that, for the PSFF, the contribution from the higher order in $1/N$ terms in $C_{\vex{n}\vex{n'}\vex{m'}\vex{m}}$ is non-negligible when $N_{\mathcal{A}}\sim O(1)$ (which is of no relevance to the current model with local Hilbert space dimension $q\rightarrow \infty$). 
The number of terms of the form $C_{\vex{n}\vex{n'}\vex{m'}\vex{m}}$ contributing to the summation in Eq.~\ref{eq:PSFF-0} is  
\begin{align}
\begin{aligned}
    \bar{N}
    =
    \begin{cases}
    NN_{\mathcal{A}},
    &
    (1) \, \vex{n}=\vex{n'},\vex{m'}=\vex{m},
    \\
    N^2/N_{\mathcal{A}}-N,
    &
    (2) \, \vex{n}=\vex{m},\vex{n'}=\vex{m'}, \vex{n} \neq \vex{n'},
    \\
    N^2-NN_{\mathcal{A}}-N^2/N_{\mathcal{A}}+N,
    &
    (3)\,
    \text{otherwise}.
    \end{cases}
\end{aligned}
\end{align}
When $N_{\mathcal{A}}\gg 1$, the numbers of total contributing terms $\bar{N}$ for categories (2) and (3) are, respectively, a factor of $O(N/N_{\mathcal{A}}^2)$ and $O(N/N_{\mathcal{A}})$ larger compared with that of categroy (1).
From the earlier result, $C_{\vex{n}\vex{n'}\vex{m'}\vex{m}}(\w)$ with indices lie in the categories (2) and (3) are of higher order in $1/N$ compared with that of category (1).
This means that we can ignore the contribution from the higher order terms in $C_{\vex{n}\vex{n'}\vex{m'}\vex{m}}$.
On the other hand, when $N_{\mathcal{A}}\sim O(1)$, the higher order in $1/N$ terms in $C_{\vex{n}\vex{n'}\vex{m'}\vex{m}}$ for categories (2) and (3) are required to obtain the leading order result for the PSFF.


\subsection{Correlation function of operators}

We now consider the dynamical correlation function of operators $O$ and $O'$:
\begin{align}
\begin{aligned}
	C_{O O'}(t)
	=
	\frac{1}{N}
	\left\langle \Tr \left( O(t) O' \right) \right\rangle 
	=
	\frac{1}{N}
	\left\langle \Tr \left( U^{\dagger}(t) O U(t) O' \right) \right\rangle,
\end{aligned}
\end{align}	
whose Fourier transform is 
\begin{align}
\begin{aligned}
	C_{OO'}(\w)
	=
	\frac{1}{N}
	\left\langle \sum_{\nu,\mu=1}^{N}
	O_{\mu\nu}O'_{\nu\mu}\delta(\w-E_{\nu}+E_{\mu})\right\rangle.
\end{aligned}
\end{align}
It is not difficult to see that the operator correlation function  $C_{OO'}(\w)$  is related to $C_{nn'm'm}(\w)$ by
\begin{align}\label{eq:CAB-0}
\begin{aligned}
	&
	C_{OO'}(\w) 
	=
	\frac{1}{N}
	\sum_{nmn'm'}
	O_{mn} O'_{n'm'}
	C_{nn'm'm}(\w).
\end{aligned}
\end{align}
For any system whose correlation function $C_{nn'm'm}(\w)$ follows the RMT prediction Eq.~\ref{eq:Cw-exact}, we have
\begin{align}\label{eq:CAB-1}
\begin{aligned}
	C_{OO'}(\w)
	=&
	\frac{ R_2(\w)-\delta(\w)}{N^2-1} 
	\left( 
	\frac{1}{N}
	\Tr(OO')
	-
	\frac{1}{N^2}\Tr(O) \Tr(O')
	\right) 
	+
	\frac{\delta(\w)}{N^2} 
	\Tr(O) \Tr(O').
\end{aligned}
\end{align}

For the current model, if we insert our leading order in $1/N$ result for the correlation function $C_{nn'm'm}(\w)$ in Eq.~\ref{eq:Cw-final}, we arrive at
\begin{align}\label{eq:CAB-2}
\begin{aligned}
C_{OO'}(\w)
\approx &
( R_2(\w)-\delta(\w)
+\frac{1}{2\pi})
\frac{1}{N^3}
\Tr(OO')
+
\left( \delta(\w)-\frac{1}{2\pi}\right) 
\frac{1}{N^2} 
\Tr(O) \Tr(O'),
\end{aligned}
\end{align}
which, for traceless operators ($\Tr O=0$, $\Tr O'=0$), reduces to
\begin{align}\label{eq:CAB-3}
\begin{aligned}
	C_{OO'}(\w)
 	\approx 
	\frac{ R_2(\w)}{N^3} 
	\Tr(OO').
\end{aligned}
\end{align}
We note that the difference between the Eq.~\ref{eq:CAB-1} and Eq.~\ref{eq:CAB-2} is due to the higher order terms in the correlation function $C_{nn'm'm}(\w)$.
These higher order terms may be non-negligible for the leading order dynamical correlation function of some operators, depending on the specific operators under consideration.

We now consider the case where the operators $O$ and $O'$ act on the same qudit.
For simplicity, we use $\vex{n}=(n_0,\vex{n}^{(L-1)})$ to label the many-body state of the entire system which consists of $L$ qudits. Here $n_0$ labels the single-particle state of the qudit on which $O$ and $O'$ operator, and the $(L-1)$-dim vector $\vex{n}^{(L-1)}$ labels the many-body state of the remaining $L-1$ qudits. Using this notation, we can see that $O_{\vex{n}\vex{m}}$ is nonzero only when $\vex{n}^{(L-1)}=\vex{m}^{(L-1)}$.
Therefore, the number of terms of the form $C_{\vex{n} \vex{n'} \vex{m'} \vex{m}}$ contributing to the summation in Eq.~\ref{eq:CAB-0} is of the order of
\begin{align}\label{eq:dc-N}
\begin{aligned}
    \bar{N}
    =
    \begin{cases}
    O(Nq),
    &
    (1) \, \vex{n}=\vex{n'},\vex{m'}=\vex{m},
    \\
    O(N^2),
    &
    (2) \, \vex{n}=\vex{m},\vex{n'}=\vex{m'}, \vex{n} \neq \vex{n'},
    \\
    O(N^2q^2),
    &
    (3)\,
    \text{otherwise}.
    \end{cases}
\end{aligned}
\end{align}
Similarly, if $O$ and $O'$ act on different qudits, we instead have
\begin{align}\label{eq:dc2-N}
\begin{aligned}
    \bar{N}
    =
    \begin{cases}
    O(N),
    &
    (1) \, \vex{n}=\vex{n'},\vex{m'}=\vex{m},
    \\
    O(N^2),
    &
    (2) \, \vex{n}=\vex{m},\vex{n'}=\vex{m'}, \vex{n} \neq \vex{n'},
    \\
    O(N^2q^2),
    &
    (3)\,
    \text{otherwise}.
    \end{cases}
\end{aligned}
\end{align}
We note that, to the second leading order,
$C_{nn'm'm}$ can be expressed as
\begin{align}\label{eq:C-2terms}
\begin{aligned}
    C_{nn'm'm}(\w)
    =\,
    c_1(\w) \delta_{nn'}\delta_{mm'}
    +
    c_2(\w) \delta_{nm}\delta_{n'm'}.
\end{aligned}
\end{align}
Here $c_{1,2}(\w)$ are functions of $\w$ (and $N$), and are independent of the specific indices $n,n',m,m'$.
This equation can be proven making use of the sigma model action $S[\tilde{Z},Z]$ (Eq.~\ref{eq:sigma}) and the fact that
the first few moments of the many-body Floquet operator $U$ are nonvanishing to the leading order only if the row/column indices of all the unitary matrices $U$ in the moment are given by the column/row indices of all the $U^{\dagger}$ matrices after a permutation~\cite{entropy}.
Specifically, 
one can apply the Wick's theorem to show that, up to the second leading order, 
$\left\langle \partial^2 S/\partial J^+_{n'n}\partial J^-_{mm'}
\right\rangle$
and
$\left\langle
(\partial S/\partial J^+_{n'n})
(\partial S/\partial J^-_{mm'})
\right\rangle
$
are nonzero only when the external Hilbert space row indices $n,m'$ are related to the column indices $n',m$ by a permutation, and they take values independent of the specific external indices.
Eq.~\ref{eq:C-2terms} then allows us to ignore the contribution from $C_{nn'm'm}(\w)$ whose indices fall into categories (2) and (3) in Eq.~\ref{eq:dc-N} or Eq.~\ref{eq:dc2-N}, for traceless local operators $O$ and $O'$.
Therefore, one can use the leading order result for $C_{nn'm'm}(\w)$ to determine the dynamical correlation function $C_{OO'}(\w)$ for the traceless local operators $O$ and $O'$ acting on the same qudit or different qudits, and the result is given by Eq.~\ref{eq:CAB-3}.


\subsection{Time evolution of the density matrix}

Using the correlation function of quasienergy eigenstates $C_{nn'm'm}(t)$, one can also study the time evolution of the density matrix $\rho(t)=U(t)\rho(0)U^{\dagger}(t)$:
\begin{align}
\begin{aligned}
    \left\langle 
    \rho_{nm}(t)
    \right\rangle
    =
    \sum_{n',m'=1}^{N}
    \rho_{n'm'}(t=0)
    C_{nn'm'm}(t),
\end{aligned}
\end{align}
where $\rho(0)$ denotes the density matrix at time $t=0$.
For the CUE, one can insert Eq.~\ref{eq:Ct-exact} into the equation above and find
\begin{align}\label{eq:rho-CUE}
\begin{aligned}
    \left\langle 
    \rho_{nm}(t)
    \right\rangle
    =
    (\rho_{th})_{nm}
    \frac{N^2-K(t)}{N^2-1}
    +
    \rho_{nm}(0)
    \frac{K(t)-1}{N^2-1},
\end{aligned}
\end{align}
where $\rho_{th}\equiv\mathbb{1}_{\mathcal{H}}/N$ is just the thermal density matrix and we have used $\Tr (\rho)=1$.
At large time $t\rightarrow \infty$, the SFF acquires the value $K(t\rightarrow \infty)=N$ and the equation above becomes
\begin{align}\label{eq:rhoinf-CUE}
\begin{aligned}
    \left\langle 
    \rho_{nm}(t\rightarrow\infty)
    \right\rangle
    =
    (\rho_{th})_{nm}
    \frac{N}{N+1}
    +
    \rho_{nm}(0)
    \frac{1}{N+1}
    \approx
    (\rho_{th})_{nm}
    +
    \rho_{nm}(0)
    \frac{1}{N}.
\end{aligned}
\end{align}
Using instead the leading order result for $C_{nn'm'm}$ derived earlier for the Floquet random quantum circuits under study, we have
\begin{align}\label{eq:rho-FRQC}
\begin{aligned}
    \left\langle 
    \rho_{nm}(t)
    \right\rangle
    =
    (\rho_{th})_{nm}
    (1-\delta_{t,0})
    +
    \rho_{nm}(0)
    \frac{K(t)-1+\delta_{t,0}}{N^2},
\end{aligned}
\end{align}
which at large time reduces to Eq.~\ref{eq:rhoinf-CUE}.
We emphasize that, the contribution from the higher order terms in $C_{nn'm'm}(t)$ to $\left\langle \rho_{nm}(t)\right\rangle$ could be as important as that from the leading order term. In this case, consideration of the higher order fluctuations in the sigma model is required to study the time evolution of the density matrix. We leave this for the future investigation.

Note that the second term in Eq.~\ref{eq:rhoinf-CUE} can be neglected when consider its contribution to the expectation value of a physical observable $O$.
Within RMT, it is straightforward to see from Eq.~\ref{eq:rho-CUE} that the expectation value of $O$ at time $t$  is 
\begin{align}\label{eq:Oinf-CUE}
\begin{aligned}
    \left\langle 
   \Tr \left( O \rho(t) \right)
    \right\rangle
    = 
    \frac{1}{N}
    \Tr(O)
    \frac{N^2-K(t)}{N^2-1}
    +
    \Tr (O\rho(0))
    \frac{K(t)-1}{N^2-1},
\end{aligned}
\end{align}
which is identical to the expression in Ref.~\cite{SFF-thermal}.
At large time $t\rightarrow \infty$, this equation reduces to
\begin{align}\label{eq:Oinf-CUE}
\begin{aligned}
    \left\langle 
   \Tr \left( O \rho(t\rightarrow\infty) \right)
    \right\rangle
    \approx 
    \frac{1}{N}
    \Tr(O)
    +
    \frac{1}{N}
    \Tr (\rho(0)O).
\end{aligned}
\end{align}
The second term in the equation above arises from $\rho_{nm}(0)/N$ in  $\left\langle 
    \rho_{nm}(t\rightarrow\infty)
    \right\rangle$ 
    (Eq.~\ref{eq:rhoinf-CUE}), and is of the order of $1/N$ smaller than the first term above, which is just the (infinite temperature) thermal average of operator $O$.

The discussion above shows that, at times larger than the plateau time (i.e., the time at which the SFF reaches the plateau $K(t)=N$), the expectation value of $O$ is given approximately by its thermal equilibrium value $\Tr(O)/N$.
The plateau time is of the order of the inverse level spacing and thus is exponentially large in system size. 
We emphasize that the relaxation time - the time at which the expectation value of observable $O$ reaches the thermal equilibrium value - is actually much smaller than the plateau time~\cite{ETH-review}. 
Let us now assume that
the density matrix evolution is indeed described by the RMT predicition Eq.~\ref{eq:rho-CUE}, and is therefore determined by the behavior of the SFF.
Note that the SFF for generic chaotic systems has a slope-ramp-plateau structure.
It drops nonuniversally from $N^2$ at $t=0$, reaches the dip, and then grows linearly (for the unitary symmetry class) until it hits the plateau at  $K(t\rightarrow \infty)=N$~\cite{Haake,Mehta}.
In this case, when the SFF takes a value $K(t) \lesssim N$, which happens at a time much earlier than the plateau time in the slope regime, the density matrix becomes approximately the thermal one $\rho_{th}=\mathbb{I}_{\mathcal{H}}/N$ in the sense that the remaining term's contribution to the expectation value of any physical observable is negligible. 
Moreover,  $\left\langle\rho(t)\right\rangle$ remains around $\rho_{th}$ as $K(t) \lesssim N$ for all remaining time.
This also suggests that the relaxation towards thermalization process is related to the early-time non-universal slope in the SFF~\cite{SFF-thermal}.
On the other hand, the RMT predicition Eq.~\ref{eq:rho-CUE} may only be valid for generic ergodic systems at late times.
It would be an interesting future direction to investigate the connection between the relaxation process and SFF, which may require consideration of higher order fluctuations as well as massive-mode fluctuations essential for the slope regime of the SFF~\cite{Altland-OS,PRB,PRR}.
Moreover, one can also estimate the statistical fluctuation of the density matrix $\rho(t)$ around the ensemble averaged result $\left\langle \rho(t) \right\rangle$ by evaluating the higher-order correlation functions of quasienergy eigenstates  in the current sigma model framework.

	\bibliography{circuit}